\documentclass[11pt,letterpaper]{article}
\usepackage[T1]{fontenc}
\usepackage[utf8]{inputenc}
\usepackage{lmodern}
\usepackage[margin=1in]{geometry}
\usepackage{amsthm,amsmath,amsfonts,amssymb}
\usepackage[authoryear,round]{natbib}
\usepackage{graphicx,enumerate,booktabs,bm}
\usepackage{algorithm,algpseudocode}
\usepackage{url}

\usepackage[colorlinks,citecolor=blue,urlcolor=blue,hypertexnames=false]{hyperref}
\newtheorem{theorem}{Theorem}

\newtheorem{proposition}{Proposition}

\newtheorem{condition}{Condition}
\theoremstyle{definition}

\title{Close Races and Similarity Bias: A Bayesian Analysis of the Ranked Ballots for the Ballon d'Or}
\author{Guanyu Hu\\
\small Department of Statistics and Probability\\
\small Michigan State University\\
\small \texttt{huguanyu@msu.edu}
\and Xiao Wang\\
\small Department of Statistics\\
\small Purdue University\\
\small \texttt{wangxiao@purdue.edu}}
\date{}

\begin{document}
\maketitle

\begin{abstract}
The Ballon d'Or ranks players whose contributions have no common unit, so its results are
often disputed. Regressions of award points describe associations but not how complete
ballots arise or how likely a different winner would be. We propose a Bayesian
mixture-of-experts model for partial rankings that learns nonlinear performance relationships,
the number of voting styles and journalists' style memberships. A shared soft regression tree
with style-specific leaf values lets styles evaluate the same season records differently on a
common basis. Region-dependent mixing probabilities and journalist--nominee similarity terms
separate geographic differences in voting styles from affinity within a style, and journalist-
and rank-specific precisions capture remaining variation. We establish identification and
consistency for the number of occupied styles under explicit conditions; simulations examine
recovery, calibration and robustness to misspecification. Applied to all 176 published ballots
from 2019 and 99 from 2024, the model compares award outcomes under changes in affinity and
jury composition. For 2019, Messi's probability of winning with a new jury from the same
countries rises from 0.49 to 0.76 when similarity terms are removed. For 2024, the direction of
this change depends on whether affinities differ by region. Regional representation changes
winner probabilities more than jury size.
\end{abstract}


\section{Introduction}\label{sec:intro}

Individual football awards turn diverse contributions to team success into a single ranking.
Goals, defending, creative play and tournament achievements have no common unit, so reasonable
observers can disagree about their value. The Ballon d'Or resolves these judgments through
journalists' ranked ballots and a scoring rule. Yet the resulting points table leaves an
important distinction hidden: a close result may reflect broad agreement with a few disputed
placements, or different groups applying different criteria. These explanations imply
different answers to a practical question: would another jury choose the same winner?
The 2019 and 2024 awards make this question concrete. Messi beat van Dijk by 686 points to
679 in 2019, despite receiving fewer first-place votes. The debate contrasted attacking
achievement with defensive contribution \citep{baldwin2019}. In 2024 Rodri beat
Vin\'icius J\'unior by 1{,}170 to 1{,}129; their different roles and club and international
achievements again offered competing grounds for judgment \citep{uefa2024}. Real Madrid's
absence from the ceremony made the disagreement especially visible \citep{hofmanfaez2024}.
Both cases raise the same issue: how much did the outcome depend on evaluation criteria,
journalist--nominee affinity and the countries represented on the jury?
We select these years as focused case studies because the results were disputed and all
published ballots are available: 176 top-five ballots in 2019 and 99 top-ten ballots in
2024. Changes to the evaluation window, criteria and jury between the awards
\citep{afp2022,tnt2022} give two contrasting voting settings, though not a causal evaluation of the reform
(nominees and performance records also changed) or a representative sample of award history.
Answering the jury question requires more than summaries of the realized vote. Point totals,
first-place counts and regional shares do not separate evaluation criteria from similarity
between journalists and nominees. Regressions of points describe conditional associations
but do not specify a distribution of ordered ballots. Without that distribution, the
probability of a different winner under another jury or scoring rule remains undefined.

A substantial literature studies similarity in award and contest voting. Shared nationality, club, league or cultural similarity is associated with voting in the FIFA award
\citep{coupe2018,johnsonmccarthy2022}, judged sports \citep{zitzewitz2006,emerson2009} and the
Eurovision Song Contest \citep{ginsburgh2008,spierdijk2009}. \citet{coupe2018} compare captains,
coaches and journalists using indicators close to ours and argue that similarity bias is most
likely to affect the Ballon d'Or winner when the leading candidates are close.
\citet{johnsonmccarthy2022} find media voters the least biased, while \citet{kopkin2022} also
find racial and religious similarity in ordered probit models of individual votes.
These studies motivate affinity terms, but the link from individual associations to an
award's winning probability requires a model for the ranked ballots themselves.

Models for rank data provide this. \citet{deng2014bard} develop Bayesian
aggregation of rank data (BARD), estimating ranker-specific reliability jointly with the
aggregate ranking and accounting for redundancy among correlated rankers.
Mixtures of experts allow ranking
preferences and mixing weights to vary across voters \citep{gormley2008aoas}; latent-class choice models identify segments with different preferences \citep{kamakura1989},
explain segment membership by covariates \citep{gupta1994}, and allow segments to differ in
ranking capability \citep{fok2012}. Bayesian Plackett--Luce mixtures and related
nonparametric models describe heterogeneous rankings and uncertainty about their groups
\citep{mollica2017,caron2014,johnson2020}, while rank-dependent precision distinguishes
disagreement at the top from less informative lower positions \citep{hausman1987,benter2008}.
Most directly related, \citet{pearce2026benter} develops Bayesian Benter inference through
positive-stable and exponential augmentation, with Gibbs updates for item worths and
Metropolis updates for dampening parameters. That model uses a common worth vector;
our focus is the interaction between rank-position noise and heterogeneous evaluation
criteria in a gated mixture.
Dyad-ranking models additionally allow preferences to depend on joint features of the ranker
and item \citep{schafer2015,schafer2018}. Within the Thurstone family,
\citet{li2022barc} develop Bayesian analysis of rank data with covariates (BARC) and
extensions for varying ranker quality and clustered opinions, including incomplete lists
and uncertainty in the aggregated ranking. Extending nonlinear ranking regression,
\citet{iacopini2026bart} introduce static ROBART and autoregressive ARROBART models,
using Bayesian additive regression trees for covariate-dependent latent scores and
allowing temporal dependence in the latter; theirs is the closest precedent for
tree-based rank models. \citet{pearce2025preferences} develop a joint
Bradley--Terry--Luce--Binomial model for rankings and ratings, using a Bayesian mixture of finite
mixtures to infer the number of preference classes and judge-specific effects to accommodate
rating behavior. Unlike independently reported ratings, Ballon d'Or points are determined
by the ballot and scoring rule. A related line of work groups the items being ranked:
\citet{santi2026btsbm} embed Bradley--Terry comparisons in a stochastic block model,
learning tied-strength groups and their uncertainty from tennis results.
\citet{santi2026plblocks} jointly group assessors and items in a Plackett--Luce latent
block model, providing a common item partition for comparing assessor preferences.

These approaches already accommodate ranker reliability, covariates, nonlinear scores and
heterogeneous preferences. We propose a Bayesian covariate-gated mixture of Benter ranking models in which all voting
styles share one soft regression tree but have their own leaf values. The model jointly learns
the tree, the number of styles and their memberships, compares the styles on a common basis
and carries their uncertainty through to the outcome of the award.
The tree captures nonlinear performance associations and interactions without prescribing
them separately for each style; shared routing and style-specific leaf values provide
regularization and a common comparison basis. An additive linear specification would require these interactions to be specified in advance.
Nominee residuals, dyadic affinity and separate journalist and rank-position precisions
account for additional ballot variation. Our methodological contributions are:

\begin{enumerate}
\item A common covariate-defined basis for comparing latent evaluation criteria
(Proposition~\ref{prop:contrast}), with regional differences entering through both style
prevalence and journalist--nominee affinity. Shared routing makes style contrasts
comparable on the same performance regions, while style-specific leaf values allow
different evaluations of similar nominee records.
\item Posterior comparisons of winning probabilities under changes in jury composition,
size, scoring rules and affinity, propagating uncertainty in the tree, styles and ballots
rather than fixing an estimated partition. These comparisons separate uncertainty about the fitted preferences from variation
across hypothetical juries.
\item Conditional identification and occupied-style consistency results
(Theorems~\ref{thm:gated}--\ref{thm:K}), together with joint transdimensional posterior
computation. The theory requires an unverified expert-identification condition and,
for consistency, a distinct-style prior restriction beyond the fitted prior.
Prior-recovery checks and simulations assess sampler implementation and finite-sample
behavior, including the difficulty of recovering weakly separated styles.
\end{enumerate}

Applied to the two awards, the framework reveals different sources of outcome uncertainty. Under the
partially pooled regional-affinity model for 2019, new journalists from the same countries
elect Messi with probability 0.49; removing affinity while holding other parameters fixed
raises this to 0.76 and increases his expected margin by 41 points (90\% interval: 17--67).
In 2024, the direction of the affinity comparison is undetermined once bonuses may vary
by region. The direction of the 2019 comparison holds across the regional-affinity specifications
we examine; the 2024 attribution does not.

Across the scenarios studied, regional representation changes winning probabilities more
than jury size does. Journalists disagree, but the ballots do not split into sharply
separated voting blocs. The comparisons remain model-based assessments of specified
changes, not causal effects of similarity bias or the jury reforms.

The remainder of the paper is organized as follows. Section~\ref{sec:data} describes the data
and sports questions. Section~\ref{sec:model} introduces the ballot model, shared tree,
priors and posterior computation. Sections~\ref{sec:q-fragility} and~\ref{sec:jury} examine the
award outcomes and the roles of affinity, voting criteria and region.
Section~\ref{sec:discussion} concludes. The Supplementary Material \citep{hu2026supp}
provides proofs, computation, simulations, additional results and model assessment.

\section{Data and Sports Questions}\label{sec:data}

\subsection{The award and its ballots}\label{sec:data-ballots}

The Ballon d'Or is awarded each year by the magazine \emph{France Football}. A jury of
journalists, one per country, ranks a published list of nominees, positions on a ballot are
converted to points, and the nominee with the most points wins. We analyze the 2019 and 2024
cycles, for which \emph{France Football} published every journalist's ballot. In 2019,
176 journalists each ranked 5 of 30
nominees, and the positions earned 6, 4, 3, 2, 1 points. In 2024, 99
journalists each ranked 10 of 30 nominees for
15, 12, 10, 8, 7, 5, 4, 3, 2, 1 points. Between the two cycles the organizer moved the evaluation window from
the calendar year to the football season, reduced the jury to about one hundred journalists, asked
voters to weight individual performance above team honors \citep{afp2022}, and removed a
player's career from the voting criteria \citep{tnt2022}. The 2024 jury was drawn from the countries in the top one hundred of the
FIFA men's ranking.

The award's stated criteria motivate the performance information collected below. For 2024,
the three criteria, in their prescribed order of importance, concern (i) individual performance,
including how decisive and impressive the player was; (ii) team performance and achievements;
and (iii) class and fair play \citep{uefa2024rules,troude2026criteria}. These are broad judgments,
not a published formula assigning numerical weights to goals, trophies or conduct. The criteria
in force in 2019 were different: individual and collective achievement during the year formed
one criterion, class including talent and fair play a second, and the player's career a third
\citep{troude2026criteria}. Thus the performance covariates address a common part of both
awards, but the 2024 ordering should not be read back into the 2019 rules.

Both races were close. In 2019 Lionel Messi
received 686 points and Virgil van Dijk 679, a gap of
7 points over 176 ballots, although van Dijk was ranked first
by more journalists (69 against
61). Plurality would therefore have elected van Dijk, whereas a
Borda count (points from the ballot length down to one) agrees with the official rule
(625 against 610). In 2024 Rodri received
1{,}170 points and Vin\'icius J\'unior 1{,}129, and Rodri
also led on first places, 49 to
35. Totals recomputed from the published ballots reproduce
the official totals except for a one-point difference between two 2019 nominees outside the top
two (Supplement S1).

\begin{figure}[!tbp]
\centering
\includegraphics[width=\textwidth]{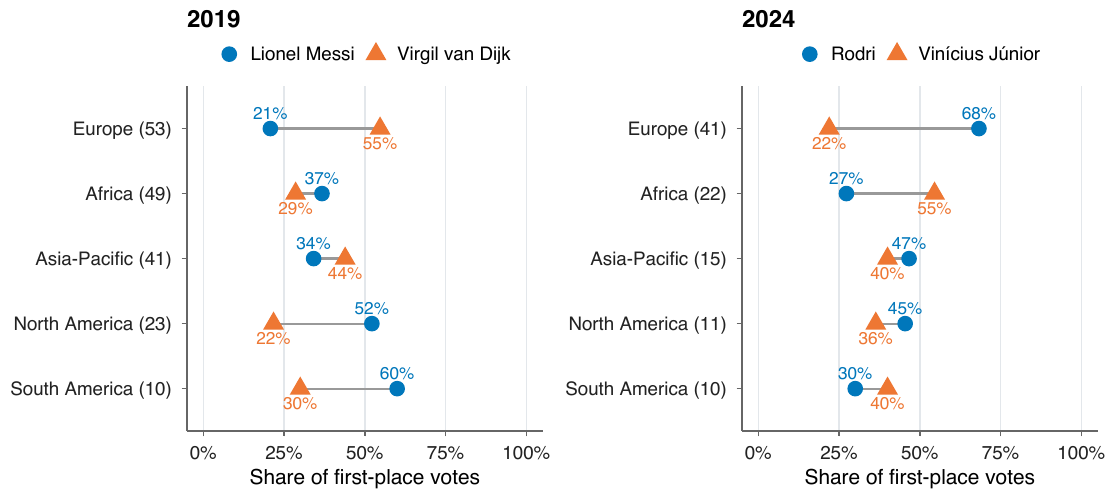}
\caption{Share of each region's first-place votes received by the official winner and runner-up.
Numbers of journalists in parentheses. Lines join the two nominees within a region.}
\label{fig:first-by-region}
\end{figure}

Figure~\ref{fig:first-by-region} shows how the first-place votes for the two leading nominees
divide by the journalist's region. In 2019, European journalists gave
55\% of their first places to van Dijk and
21\% to Messi, whereas journalists from North and Central America
and from South America gave Messi 52\% and
60\% of theirs. In 2024 Europe favored Rodri
(68\% of European first places) and Africa favored Vin\'icius
(55\%). These raw splits combine at least three sources of
variation: regions may weigh the season record differently, journalists may favor players with
whom they share a nation, confederation or league, and a region with ten journalists produces
noisy shares. Separating these sources is the purpose of the model in Section~\ref{sec:model}.

\subsection{Nominee information}\label{sec:data-nominees}

We choose covariates to describe the observable performance and achievement dimensions of these
criteria while keeping their interpretation comparable across nominees. A list limited to goals
and trophies would favor particular types of contribution before the ballots are analyzed:
scoring is central to an attacker's record, but ball progression, chance creation and defensive
work help describe the records of midfielders and defenders. Conversely, individual event
statistics alone would omit the collective success that voters are explicitly asked to consider.
The resulting four groups organize the fitted performance surface for interpretation; they are
not official criteria (Supplement Table~S1).

The \emph{individual} group describes the first performance dimension through league goals and
assists, non-penalty expected goals and expected assisted goals, key passes, progressive carries,
passes and passes received, successful take-ons, and tackles and interceptions. The rate
measures are recorded per 90 minutes, alongside the season totals for goals and assists. These variables cover finishing, chance creation, progression and ball-winning. An indicator for a top-five
Champions League scorer adds information about scoring prominence in European competition.
It is a coarse measure of individual distinction. The public scorer tables list only players above a goal threshold, so an indicator
of appearing among the leaders avoids treating an unlisted player's goal count as zero.

The \emph{club} group addresses collective achievement through domestic league finish, the main
domestic cup result and the stage reached in European competition. The \emph{international}
group extends the record to national-team tournaments using tournament outcome and goals.
The former describes collective success and the latter individual contribution, so this group
spans the first two official performance dimensions; it is separated from club results because
club and national-team achievements need not be valued in the same way. Tournament outcomes
record the success of a nominee's team, not the share of that success attributable to the
nominee. A nominee's club is the club represented during the evaluation window, not the club
at the ceremony.

The \emph{role} group contains position, age and an indicator of playing outside the five major
European leagues. These are contextual variables, not official award criteria. Position lets
the model relate performance measures to the kind of contribution a player makes, including
interactions between role and the season record in the shared tree. Age permits differences
associated with career stage without treating age as a measurement of career achievement.
The league indicator records a different competitive and data-coverage context, particularly
for the five 2019 nominees whose league measures require imputation.

The third 2024 criterion, class and fair play, is not directly measured in the harmonized data.
Nor do we observe a comparable career-achievement history for the 2019 nominees. The covariates therefore do not cover all the official criteria. The nominee effect in
Section~\ref{sec:model} absorbs these omissions, other aspects of performance and model error,
but cannot identify a separate fair-play or career effect.

The 2024 window is the 2023--24 club season with the 2024 European Championship, the 2024 Copa
Am\'erica and the 2023 Africa Cup of Nations, which was played in January and February 2024.
The 2019 award covered the calendar year, but season statistics are compiled by season, so we
use the 2018--19 club season, the tournaments of summer 2019 and the Asian Cup of January 2019.
Club form between August and November 2019, which voters had seen, is therefore not in the
covariates; Section~\ref{sec:model} introduces a nominee-specific residual that absorbs this and other appeal the covariates do not record.

Supplement S1 lists the corrections made to the supplied nominee file, each with
its source. In 2019, 5 nominees played for Ajax or Benfica, outside the
coverage of the league measures; their league measures are imputed by the mean over nominees in
the same position, and the outside-the-five-leagues indicator lets the model shift their utility.
This indicator is constant, and therefore dropped, in 2024.

\subsection{Journalist information}\label{sec:data-journalists}

A journalist is identified by country, and the country's confederation gives the region:
Europe (UEFA), Africa (CAF), Asia--Pacific (AFC and OFC; the latter has
4 journalists in 2019 and 1 in 2024), North and Central
America (CONCACAF; ``North America'' in tables and figures) and South America (CONMEBOL). The 2019 jury comprises 53
European, 49 African, 41 Asia--Pacific,
23 North and Central American and 10 South American
journalists. The 2024 jury is smaller and more European: 41,
22, 15, 11 and
10 in the same order, so Europe's share rises from
30\% to 41\%.

For each journalist--nominee pair we record three affinity indicators: the journalist's country
is the nominee's nationality (\emph{same nation}), the two belong to the same confederation
(\emph{same confederation}), and the nominee's club plays in the journalist's country
(\emph{same league country}). Affinity is visible in the raw ballots. In 2019 a journalist
ranked a compatriot nominee in 50.0\% of the 30
compatriot pairs, against 16.5\% for all other pairs; in 2024 the
corresponding rates are 56.7\% and 33.1\% (the base
rate is higher because the 2024 ballot is twice as long), and a nominee playing in the journalist's
country is ranked in 50.0\% of such pairs. Such contrasts do not separate
affinity from quality: a country whose compatriot is a strong candidate would rank that nominee
often without any preference for compatriots. The model in Section~\ref{sec:model} estimates affinity
conditional on each nominee's utility.

Unlike the performance covariates, the region variables and affinity indicators describe who
votes and the journalist's relationship to a nominee. They let us ask whether geographic
patterns remain after accounting for the recorded performance and achievement dimensions of
the award criteria, subject to the omissions described above.

\subsection{Sports questions}\label{sec:questions}

The ballots allow five questions about how an international jury reaches a collective ranking.
Each question motivates a component of the model in Section~\ref{sec:model} and is answered in
the section named below.

\begin{enumerate}[Q1.]
\item \emph{How fragile were the 2019 and 2024 results, and what decided them?} We call a result fragile if a new jury drawn from the same countries would elect a different
winner with non-negligible probability. We ask how often the official winner would win
if the same journalists, or new journalists from the same countries, voted again, and how that probability changes when journalist--nominee affinity is
removed, when the other cycle's jury votes, when only European or only non-European journalists
vote, and when plurality or Borda scoring replaces the official rule. The organizer chooses the
jury's countries and the scoring rule, so these are the levers it controls
(Section~\ref{sec:q-fragility}).

\item \emph{What does a ballot reward?} We ask how strongly journalists favor nominees of their
own nation, confederation or league country once quality is accounted for, how steeply attention
declines down a ballot, and how much of a nominee's standing each covariate group explains, and how much is left to
unrecorded factors such as reputation
(Section~\ref{sec:q-ballot}).

\item \emph{Is there one electorate or several voting styles?} If journalists differ in what they
reward, the aggregate ranking averages over distinct criteria. The answer requires inferring the
number of styles from the ballots rather than fixing it (Section~\ref{sec:q-styles}).

\item \emph{Does a journalist's region predict the criteria the journalist applies?} If it does,
the jury's regional composition shapes the result through criteria and not only through affinity
(Section~\ref{sec:q-region}).

\item \emph{Do these patterns recur across the two cycles?} The 2022 changes to the jury and to
the evaluation window make the two cycles a natural comparison, but different nominees and a
single change of rules do not permit a causal statement about the reform; the comparison is
descriptive (Section~\ref{sec:q-cycles}).
\end{enumerate}

\section{Model}\label{sec:model}

\begin{figure}[!tbp]
\centering
\includegraphics[width=\textwidth]{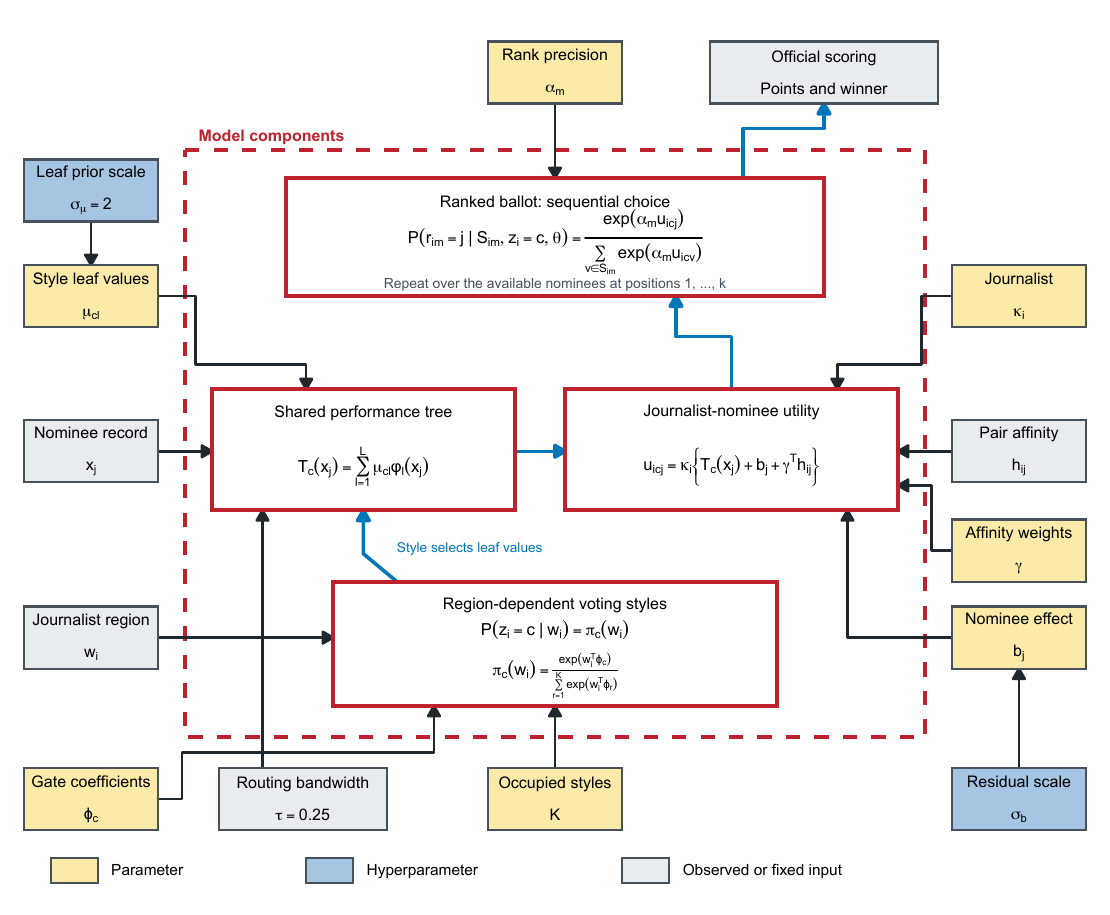}
\caption{Hierarchical representation of the ranking model. The red-bordered components show
the region-dependent voting styles, shared performance tree, journalist--nominee utility and
sequential ballot choices. Here $T_c(x_j)=T(x_j;{\boldsymbol{\mu}}_c)$ and
$\mathcal S_{im}$ is the set of nominees available at position $m$. Blue arrows connect model
components and the scoring of the resulting ballot; black arrows connect parameters and inputs
to their components. Yellow boxes denote parameters, blue boxes hyperparameters and gray boxes
observed or fixed quantities. The leaf scale is fixed at $\sigma_\mu=2$ and the routing
bandwidth at $\tau=0.25$; the residual scale $\sigma_b$ is learned.}
\label{fig:hierarchy}
\end{figure}

We fit the model separately to each cycle. Journalists are indexed by $i=1,\ldots,n$, nominees
by $j=1,\ldots,q$, and $x_j\in{\mathbb{R}}^p$ denotes nominee $j$'s covariates, standardized within
the cycle. Journalist $i$ reports the ballot $R_i=(r_{i1},\ldots,r_{ik})$, where $r_{im}$ is the
nominee placed in position $m$; the remaining $q-k$ nominees are unranked and unordered.
Figure~\ref{fig:hierarchy} shows how the parts of the model generate a ballot.

\subsection{Ballots as sequential choices}\label{sec:model-likelihood}

Given a voting style $z_i=c$, we model the ballot as a sequence of choices from the nominees
still available, $\mathcal S_{im}=\{1,\ldots,q\}\setminus\{r_{i1},\ldots,r_{i,m-1}\}$:
\begin{equation}
P(R_i\mid z_i=c,\theta)=\prod_{m=1}^{k}
\frac{\exp\{\alpha_m\,u_{icr_{im}}\}}{\sum_{j\in\mathcal S_{im}}\exp\{\alpha_m\,u_{icj}\}},
\qquad \alpha_1=1 .
\label{eq:likelihood}
\end{equation}
Setting $\alpha_m\equiv1$ gives the top-$k$ Plackett--Luce model
\citep{plackett1975,luce1959}. Unranked nominees enter the choice-set denominators but
contribute no relative ordering. Following \citet{benter2008} and \citet{gormley2008aoas},
positive rank-position precisions allow later choices to be less ($\alpha_m<1$) or more
($\alpha_m>1$) concentrated than the first. They are shared across journalists and styles,
with no monotonicity constraint. Maximum-likelihood fits show declining precision, and
the Bayesian information criterion favors a single Benter ranking over every fitted
Plackett--Luce mixture (Supplement S6), so we model positional noise separately from style heterogeneity.

The utility of nominee $j$ for journalist $i$ under style $c$ is
\begin{equation}
u_{icj}=\kappa_i\bigl\{T(x_j;{\boldsymbol{\mu}}_c)+b_j+{\boldsymbol{\gamma}}^\top h_{ij}\bigr\},
\label{eq:utility}
\end{equation}
combining style-specific performance, a shared nominee residual and journalist--nominee
affinity. The journalist precision $\kappa_i>0$ controls concentration across the whole
ballot, whereas $\alpha_m$ varies by position. Allowing low precision for atypical ballots
reduces pressure to create small styles solely to accommodate noise
\citep{cai2021,johnson2020}. For two available nominees $j,j'$ at position $m$, the
conditional choice $J_{im}$ satisfies
\[
\begin{aligned}
\log\frac{P(J_{im}=j\mid\mathcal S_{im},z_i=c,\theta)}
              {P(J_{im}=j'\mid\mathcal S_{im},z_i=c,\theta)}
&=\alpha_m\kappa_i\bigl[
T(x_j;{\boldsymbol{\mu}}_c)-T(x_{j'};{\boldsymbol{\mu}}_c)\\
&\qquad{}+b_j-b_{j'}+{\boldsymbol{\gamma}}^\top(h_{ij}-h_{ij'})\bigr].
\end{aligned}
\]
These are choice-set odds, not marginal finishing-order probabilities. Ballots are
conditionally independent given styles and parameters before occupancy conditioning
(Section~\ref{sec:model-prior}).

\subsection{A shared soft tree with style-specific leaves}\label{sec:model-tree}

The covariates describe individual performance, club and international achievements, and
playing role (Section~\ref{sec:data-nominees}); all retained variables, including binary
indicators, are standardized within cycle. A linear specification $x_j^\top\beta_c$
would impose additive, constant slopes unless interactions were specified explicitly.
For example, the value of additional goals could depend on position or club success.
Because each cycle contains only 30 distinct nominee records, we use one regularized
nonlinear basis shared across styles rather than separate high-dimensional functions.

Specifically, all styles share a soft binary tree and differ in the values they attach to its
leaves. A binary tree with internal nodes $v$ splits on covariate $s_v$ at
threshold $t_v$, but the split is soft: nominee $j$ goes left at $v$ with probability
$\psi_v(x_j)=\operatorname{expit}\{(t_v-x_{js_v})/\tau\}$, where $\operatorname{expit}(a)=1/(1+e^{-a})$ is the logistic function and
$\tau$ a fixed bandwidth \citep{irsoy2012,linero2018}. The probability that $x_j$ reaches leaf
$l$ is the product of the branch probabilities along the path,
$\phi_l(x_j)=\prod_{v\in\mathrm{path}(l)}\psi_v(x_j)^{\mathbb 1[l\in\mathrm{left}(v)]}
\{1-\psi_v(x_j)\}^{\mathbb 1[l\in\mathrm{right}(v)]}$, and style $c$ assigns
\begin{equation}
T(x_j;{\boldsymbol{\mu}}_c)=\sum_{l=1}^{L}\mu_{cl}\,\phi_l(x_j).
\label{eq:tree}
\end{equation}
The routing weights are nonnegative and sum to one. All styles share the topology,
split variables and thresholds but have distinct leaf values, giving $KL$ coefficients
on a common basis. The tree prior favors small bases; soft routing smooths threshold
effects and approaches hard routing as $\tau\to0$. We fix $\tau=0.25$, examine alternatives
in sensitivity analyses, and estimate the remaining tree parameters jointly.

Style contrasts satisfy
$T(x_j;{\boldsymbol{\mu}}_c)-T(x_j;{\boldsymbol{\mu}}_r)
=\sum_l(\mu_{cl}-\mu_{rl})\phi_l(x_j)$, permitting comparisons on the same basis while
propagating tree and label uncertainty. Splits describe associations, not independently
identified or causal effects: correlated covariates may substitute for one another.
Unlike the item blocks of \citet{santi2026plblocks}, leaves are covariate-defined and
membership is soft, without imposing equal nominee utilities. Sharing a topology trades some flexibility for regularization and comparability. Supplement S8
compares it with free nominee utilities but not with a fitted linear model.

The shared residual $b_j\sim N(0,\sigma_b^2)$ absorbs unrecorded standing, including
reputation, omitted performance and model error; in 2019 it can also reflect career and
post-season information (Section~\ref{sec:data-nominees}). A common residual can accommodate a nominee rated above the recorded performance
without creating another style, and cancels from between-style contrasts.

\subsection{Journalist--nominee affinity}\label{sec:model-affinity}

The vector $h_{ij}\in\{0,1\}^3$ indicates shared nation, confederation and league country
(Section~\ref{sec:data-journalists}). Bonuses are additive; since shared nationality implies
shared confederation, $\gamma_{\mathrm{nation}}$ is the additional compatriot bonus.
Following studies of football awards \citep{coupe2018,johnsonmccarthy2022} and judged
contests \citep{zitzewitz2006,emerson2009,ginsburgh2008,spierdijk2009}, we call these
associations similarity bias. This is descriptive: the model cannot separate preference
from familiarity \citep{li2017} or geographically aligned criteria. Including affinity
helps distinguish these associations from regional differences in voting styles.

The base coefficients are shared across styles and regions. Two variants relax regional
pooling of the confederation bonus. One assigns region $\rho$ a coefficient $\gamma_{\mathrm{conf},\rho}$ with independent $N(0,1.0^2)$ priors. The other pools them
partially, $\gamma_{\mathrm{conf},\rho}=\bar\gamma+\zeta_\rho$ with $\bar\gamma\sim N(0,1.0^2)$,
$\zeta_\rho\sim N(0,\tau_\gamma^2)$ and $\tau_\gamma\sim N^+(0,1.0^2)$.
Bonuses for regions without a nominee from their confederation (North America;
Asia--Pacific in 2024) do not enter the likelihood (Supplement S2.2). Because the partially pooled variant predicts better in 2019 (Section~\ref{sec:q1-affinity}),
we report outcome attribution under both specifications; other analyses use the common bonus.

\subsection{Covariate-gated voting styles}\label{sec:model-gate}

Journalist $i$ belongs to style $z_i\in\{1,\ldots,K\}$ with probability
\begin{equation}
P(z_i=c\mid w_i)=\pi_c(w_i)=
\frac{\exp(w_i^\top\varphi_c)}{\sum_{r=1}^{K}\exp(w_i^\top\varphi_r)},
\label{eq:gate}
\end{equation}
where $w_i$ contains an intercept and region indicators, with Europe as baseline.
Journalists in a region share mixing probabilities, not necessarily styles. The gate
changes the distribution of criteria; affinity changes pair-specific utility. The data separate the two only partly. This mixture-of-experts construction
\citep{jacobs1991,gormley2008aoas} parallels latent-class choice models \citep{kamakura1989}, in particular those with
covariate-dependent membership \citep{gupta1994}. Because common shifts of $\varphi_c$ leave the gate
unchanged, we use exchangeable priors and report
$\delta_c=\varphi_c-K^{-1}\sum_r\varphi_r$.

\subsection{The number of styles and the prior}\label{sec:model-prior}

The number of styles is unknown. We let $K$ count occupied styles: the model is conditioned on
the event $\mathcal O_K$ that each of the $K$ styles has at least one journalist, so the joint
posterior is
\begin{equation}
p(K,\lambda,\theta,z\mid R)\;\propto\;p(K\mid\lambda)\,p(\lambda)\,p(\theta\mid K)
\prod_{i=1}^{n}\pi_{z_i}(w_i)\,P(R_i\mid z_i,\theta)\;\mathbb 1\{z\in\mathcal O_K\},
\label{eq:posterior}
\end{equation}
with a truncated shifted Poisson prior $p(K\mid\lambda)\propto\lambda^{K-1}/(K-1)!$ on
$\{1,\ldots,K_{\max}\}$ and $\lambda\sim\mathrm{Gamma}(1.0,1.0)$, a mixture of
finite mixtures in the sense of \citet{miller2018} and \citet{fruhwirth2021}.
\citet{pearce2025preferences} use an MFM for ranking and rating preferences; our gate and
occupancy restriction give a different joint target. Occupancy conditioning multiplies
the prior mass of $K$ by $P(\mathcal O_K\mid K)$, equal to one at $K=1$ and 0.96 and 0.75
at $K_{\max}$ in 2019 and 2024 (Supplement S3.1). We display this induced prior and verify
its recovery without the likelihood (Supplement S4). We choose $K_{\max}=3$ and $5$,
respectively, from the largest $K$ outside the non-identifiable region $k\le2K-2$ for
free-utility Plackett--Luce mixtures \citep{zhao2019}. This guides the design but does not
guarantee identification of the gated Benter model.

Under the tree prior of \citet{chipman1998}, a node at depth $d$ splits with
probability $0.95(1+d)^{-2}$; its split variable is uniform over the retained covariates and its
threshold has a standard normal prior on the standardized scale. Conditional on the tree,
$\mu_{cl}\sim N(0,2.0^2)$ independently across styles and leaves. The unexplained nominee
effects satisfy $b_j\mid\sigma_b\sim N(0,\sigma_b^2)$ with
$\sigma_b\sim N^+(0,1.0^2)$, and the three common affinity coefficients have independent
$N(0,1.0^2)$ priors. Each five-dimensional gate coefficient has prior
$\varphi_c\sim N_5(0,1.50^2 I_5)$. We use
$\kappa_i\sim\mathrm{Gamma}(3.0,3.0)$, in shape--rate notation, so its mean is one, and
$\log\alpha_m\sim N(0,2.0^2)$ independently for $m\geq2$, with $\alpha_1$ fixed at one.
These priors precede occupancy conditioning; regional-affinity variants use
Section~\ref{sec:model-affinity}'s hierarchy. Supplement Table~S3 collects the priors, and
Supplement S7 examines hyperparameter sensitivity.

Supplement S8 compares the full model with reduced models: a single
Plackett--Luce ranking with free nominee utilities (M1); one style with the tree, residual,
affinity and both precisions (M2); styles with a gate that ignores region (M3); gated styles with
free utilities in place of the tree (M4); and the full model without, in turn, the residual (M5b),
the affinity terms (M5n), the journalist precisions (M5u) or the rank-position precisions (M5p).

\subsection{Identification and posterior learning}\label{sec:theory}

Only centered utilities enter the likelihood (Supplement Lemma~S1); their scale is fixed
by the precision distribution. We report centered utilities, shared-tree style contrasts
and winning probabilities. Individual leaf values and the tree--residual decomposition
remain prior-dependent.

\emph{Theoretical setup.} Write $\Phi_{jl}=\phi_l(x_j)$ for the shared routing matrix,
$u_c=\Phi{\boldsymbol{\mu}}_c+b$ for the baseline utility vector of style $c$, and
$C_q=I_q-q^{-1}{\mathbf{1}}_q{\mathbf{1}}_q^\top$, so $\tilde u_c=C_qu_c$ is centered.
The centered gate coefficients are $\delta_c=\varphi_c-K^{-1}\sum_r\varphi_r$.

\begin{proposition}[Style contrasts through the shared tree]\label{prop:contrast}
For styles $c$ and $r$,
\[
\tilde u_c-\tilde u_r=C_q\Phi({\boldsymbol{\mu}}_c-{\boldsymbol{\mu}}_r),
\]
which does not involve $b$. If $\Phi$ has full column rank, then $\tilde u_c=\tilde u_r$ if and
only if ${\boldsymbol{\mu}}_c-{\boldsymbol{\mu}}_r$ is a multiple of ${\mathbf{1}}_L$. Thus, for a
given tree, the leaf-value contrasts are determined up to a common shift by the utility
contrasts. Each function $x\mapsto T(x;{\boldsymbol{\mu}}_c)$ is infinitely differentiable.
If $D_s$ is the largest number of splits on covariate $s$ along any root-to-leaf path, then
\[
\left|\frac{\partial T(x;{\boldsymbol{\mu}}_c)}{\partial x_s}\right|
\leq\frac{D_s}{4\tau}\left(\max_l\mu_{cl}-\min_l\mu_{cl}\right).
\]
\end{proposition}

For identification, let $Q$ be the parameter-independent distribution of the country-derived
profiles $(w,h)$, with finite support, where $h$ collects the $q$ nominees' affinity vectors.
Let $f_\alpha(R\mid v)$ denote the ranking probability in \eqref{eq:likelihood} with utilities
$v$, and integrate a new journalist's precision against its fixed prior law:
\[
g_0(R\mid v;\alpha)=\int_0^\infty f_\alpha(R\mid\kappa v)
\,\mathrm{Gamma}(d\kappa;3.0,3.0).
\]
The conditional ballot law is
\[
p_\vartheta(R\mid w,h)=\sum_{c=1}^K\pi_c(w)
g_0(R\mid u_c+\gamma^\top h;\alpha).
\]
Here $\vartheta$ collects the centered utilities, centered gate coefficients, affinity
coefficients and rank-position precisions, rather than a separately identified tree and
residual decomposition. We require the following two conditions.

\begin{condition}[Identification of the experts]\label{cond:experts}
For $K,K'\leq K_{\max}$, suppose
\[
\sum_{c=1}^K a_c\,g_0(\cdot\mid v_c;\alpha)
=\sum_{r=1}^{K'}a'_r\,g_0(\cdot\mid v'_r;\alpha')
\]
with positive weights and distinct centered vectors within each mixture. Then $K=K'$,
$\alpha=\alpha'$, and $\{(a_c,C_qv_c)\}_{c\leq K}=
\{(a'_r,C_qv'_r)\}_{r\leq K'}$ after a permutation.
\end{condition}

\begin{condition}[Affinity variation]\label{cond:affinity}
The only $\chi\in{\mathbb{R}}^3$ satisfying $C_q\chi^\top(h-h')=0$ for every $h,h'$ in the support of $Q$ is $\chi=0$.
\end{condition}

\begin{theorem}[Identification of the gated model]\label{thm:gated}
Suppose Conditions~\ref{cond:experts} and~\ref{cond:affinity} hold and the region vectors $w$ in
the support of $Q$ span ${\mathbb{R}}^5$. Let $\vartheta$ and $\vartheta'$ have
$K,K'\leq K_{\max}$ styles, each with distinct centered utilities. If
$p_\vartheta(\cdot\mid w,h)=p_{\vartheta'}(\cdot\mid w,h)$ on the support of $Q$, then $K=K'$,
$(\gamma,\alpha)=(\gamma',\alpha')$, and
\[
\{(\tilde u_c,\delta_c)\}_{c\leq K}
=\{(\tilde u'_c,\delta'_c)\}_{c\leq K}
\]
after a common permutation.
\end{theorem}

Condition~\ref{cond:experts} is a global expert-mixture identification assumption, which we
cannot verify for the precision-mixed Benter model; Theorem~\ref{thm:gated} is therefore
conditional. Condition~\ref{cond:affinity} holds for the three common affinity indicators in
both cycles (rank 3), and the gate design spans its five-dimensional space because all five
regions are represented. The regional-bonus variants require deleting coefficients whose
indicators are identically zero before applying the rank condition (Supplement S2.2).
The theorem identifies the listed utility and gate quantities, not the tree, its individual
leaf values or the division of standing between the tree and $b$.

Let $\Theta=\bigcup_{K\leq K_{\max}}\Theta_K$, where $\Theta_K$ contains these identified
parameters with distinct centered style utilities, modulo label permutations. Write $\Pi$ for
the induced unconditioned prior, $D_n=\{(R_i,w_i,h_i)\}_{i=1}^n$, and $\Pi_{\mathcal O}$ for
the occupancy-conditioned posterior of \eqref{eq:posterior}.

\begin{theorem}[Consistency for the number of styles]\label{thm:K}
Suppose $(R_i,w_i,h_i)$ are independent draws from
$p_{\vartheta_0}(R\mid w,h)\,Q(dw,dh)$, the assumptions of Theorem~\ref{thm:gated} hold, and
the prior gives probability one to $\Theta$ and positive probability to every
$K\leq K_{\max}$. Then, for $\Pi$-almost every $\vartheta_0$ with $K_0$ styles,
\[
\Pi_{\mathcal O}(K=K_0\mid D_n)\longrightarrow1
\]
almost surely. The posterior mode of $K$ equals $K_0$ for all sufficiently large $n$, almost
surely.
\end{theorem}

The fitted prior permits an unsplit root, for which styles have identical centered
utilities. Theorem~\ref{thm:K} instead applies after conditioning on a split root, ensuring
distinct utilities almost surely. This conditioning changes posterior event probabilities
by at most the posterior probability of a single-leaf tree, 0.006 in 2019 and below 0.0005 in 2024 (Supplement S2), but does not establish consistency under the fitted unconditioned
prior. The result has no rate, holds only for prior-almost every true parameter, and
concerns a hypothetical journalist population. We treat $K$ descriptively for these juries.

Proofs of Proposition~\ref{prop:contrast} and Theorems~\ref{thm:gated}--\ref{thm:K} are in
Supplement S2, together with the invariance lemma and the local-identifiability result.
Supplement S6.3 checks local identifiability at posterior draws. The predictive winning
probabilities remain the primary outcome-level summaries; their sensitivity to $K_{\max}$,
chains and model classes is assessed in Section~\ref{sec:q-styles} and Supplement S8.

\subsection{Posterior computation}\label{sec:sampling}

We sample \eqref{eq:posterior} by Metropolis-within-Gibbs, retaining the leaf values and
gate coefficients explicitly. Allocation, dimension-changing, Hamiltonian and tree updates
jointly explore the occupied styles and performance basis. Supplement S3 gives full
proposal densities, Jacobians and acceptance ratios.

For the current tree $\mathcal T$ and continuous parameters $\theta$, define
\[
\ell_i(c)=\log P(R_i\mid z_i=c,\theta),
\qquad
W_{ic}=\pi_c(w_i)\exp\{\ell_i(c)\}.
\]
If journalist $i$ belongs to a style with at least two members, the allocation update is
\begin{equation}
P(z_i=c\mid\mathrm{rest})
=\frac{W_{ic}}{\sum_{r=1}^{K}W_{ir}},
\qquad c=1,\ldots,K.
\label{eq:allocation-update}
\end{equation}
Singleton allocations remain fixed in this block. Birth--death moves add or remove
singleton styles; split--merge moves reallocate larger groups. Births draw new parameters
from their priors, while splits use opposite perturbations. Acceptance ratios include
the change in $K$, proposal probabilities and all journalists' softmax normalizers.
Splits creating an empty style are rejected.

For the continuous updates, write $b_j=\sigma_b\xi_j$ with $\xi_j\sim N(0,1)$ and let
\[
\eta=({\boldsymbol\mu},\xi,\log\sigma_b,\gamma,
       \log\alpha_2,\ldots,\log\alpha_k).
\]
Conditional on $(K,\mathcal T,z,\varphi,\kappa)$, the log target for $\eta$ is
\begin{equation}
\log p(\eta\mid\mathrm{rest})
=\sum_{i=1}^{n}\ell_i(z_i)+\log p({\boldsymbol\mu},\xi,\sigma_b,\gamma,\alpha
\mid K,\mathcal T)
+\log\sigma_b+\sum_{m=2}^{k}\log\alpha_m+C,
\label{eq:continuous-target}
\end{equation}
where $C$ is constant in $\eta$ and the last two terms are transformation Jacobians.
We update this block by Hamiltonian Monte Carlo using analytic likelihood gradients.
The gate coefficients have a separate Hamiltonian update with log target
\[
\log p(\varphi\mid\mathrm{rest})
=\sum_{i=1}^{n}\log\pi_{z_i}(w_i)
-\frac{1}{2(1.50)^2}\sum_{c=1}^{K}\|\varphi_c\|^2+C.
\]
Each journalist precision is updated on the log scale. With
$\kappa_i'=\kappa_i\exp(0.5\epsilon_i)$ and $\epsilon_i\sim N(0,1)$, the acceptance
probability is
\begin{equation}
\min\left\{1,
\exp\{\ell_i'(z_i)-\ell_i(z_i)\}
\frac{g_\kappa(\kappa_i')\kappa_i'}{g_\kappa(\kappa_i)\kappa_i}
\right\},
\label{eq:precision-update}
\end{equation}
where $g_\kappa$ is the $\mathrm{Gamma}(3,3)$ prior density and $\ell_i'$ uses the
proposed precision. Common-shift and scale-ridge moves additionally explore the
likelihood-invariant directions described in Supplement S2.

Threshold, change and grow--prune moves update the tree. Growing replaces a leaf by two
with oppositely perturbed values in each style; pruning reverses the move. Acceptance
ratios include tree and leaf priors, all ballot likelihoods and the Jacobian.
A random walk on $\log\lambda$ targets
\[
p(\lambda\mid\mathrm{rest})\propto
p(\lambda)\,p(K\mid\lambda),
\]
including the normalizing constant of the truncated shifted Poisson distribution.

For 2024, we use parallel tempering with 9 replicas and ballot-likelihood powers
$1=\nu_1>\cdots>\nu_9=0.48$. The gate and priors are not tempered. Write
$\ell(\theta,z)=\sum_i\ell_i(z_i)$. After each iteration, a uniformly selected adjacent
pair $(t,t+1)$ proposes to exchange its complete states with probability
\begin{equation}
\min\left\{1,\exp\left[(\nu_t-\nu_{t+1})
\{\ell(\theta_{t+1},z_{t+1})-\ell(\theta_t,z_t)\}\right]\right\}.
\label{eq:tempering-update}
\end{equation}
Only the untempered replica supplies posterior draws. Warm-up adaptation is frozen
before draws are recorded. The additional split--merge and tree attempts, mass-matrix
adaptation and residual-scale interweaving used for the reported 2024 fits are specified
in Supplement S3.

Algorithm~\ref{alg:posterior} summarizes one chain; in the tempered analysis its update
blocks are applied to each replica at its own likelihood power. We use 8 independent
chains with 600{,}000 post-warm-up iterations after 40{,}000 warm-up iterations for 2019,
and 300{,}000 after 15{,}000 for 2024. Sampler validation is in Supplement S4, and
convergence diagnostics and Monte Carlo error assessments are in Supplement S6.1.
The latter limit the 2024 style interpretation to qualitative summaries and its
auxiliary fits to indicative comparisons.

\begin{algorithm}[t]
\caption{Posterior sampling for the shared-tree gated ranking model}
\label{alg:posterior}
\begin{algorithmic}[1]
\State Initialize an occupied allocation $z$, $K$, $\mathcal T$, $\lambda$ and $\theta$.
\For{$s=1,\ldots,S_{\mathrm{total}}$ iterations}
    \For{each journalist $i$ in a nonsingleton style}
        \State Sample $z_i$ using \eqref{eq:allocation-update}.
    \EndFor
    \State Propose birth--death and split--merge updates of $K$, $z$,
    \Statex \hspace{\algorithmicindent} the leaf values and the gate coefficients.
    \State Update $\eta$ by Hamiltonian Monte Carlo targeting \eqref{eq:continuous-target}.
    \State Update the gate coefficients by a separate Hamiltonian step.
    \State Update each $\kappa_i$ using \eqref{eq:precision-update}.
    \State Apply common-shift and scale-ridge updates.
    \State Update $\mathcal T$ by threshold, change and grow--prune proposals.
    \State Update $\log\lambda$ by random-walk Metropolis.
    \If{parallel tempering is used}
        \State Propose an adjacent-replica exchange using \eqref{eq:tempering-update}.
    \EndIf
    \If{warm-up is complete and this iteration is selected for storage}
        \State Store $(K,z,\mathcal T,\lambda,\theta)$ from the untempered replica.
    \EndIf
\EndFor
\end{algorithmic}
\end{algorithm}

\section{Fragility of the 2019 and 2024 Results}\label{sec:q-fragility}

Sections~\ref{sec:q-fragility} and~\ref{sec:jury} take the questions of
Section~\ref{sec:questions} in turn; Supplement S8 assesses the model. All
summaries come from the full model fitted separately to each cycle; Table~\ref{tab:scenarios} and
Section~\ref{sec:q1-affinity} also use its variant with partially pooled regional
bonuses (Section~\ref{sec:model-affinity}), and Table~\ref{tab:shift} further specifications. We write \emph{pooled} for partially pooled below.
This section answers Q1.

\subsection{Counterfactual juries}\label{sec:q1-scenarios}

A scenario fixes a posterior draw $\theta$ and simulates a jury: a list of journalists, each with a
region vector and affinity indicators, who draw their styles from the gate and their precisions
from the prior (in a re-vote the observed journalists keep those of the draw) and cast one top-$k$
ballot each. Unranked nominees receive no points, and the winner is the nominee with the most
official points among all 30, with ties broken by first places and then at random.
Table~\ref{tab:scenarios} reports, with a common and with partially pooled regional bonuses, the
posterior mean of $P(\mathrm{win}\mid\theta)$, the probability that the official winner wins, which
is the posterior predictive probability, and its 90\% interval across draws. The posterior predictive
probability averages over the parameters and over the randomness of a new jury. The interval shows
the uncertainty about the parameters alone, and the finite number of simulated juries adds the Monte
Carlo error bounded in the caption. The
juries are the observed jury, with the observed or new journalists; the same with $\gamma=0$ and
every other parameter unchanged; this cycle's journalists resampled with replacement within region
to the other cycle's regional shares at this cycle's size, or to the other cycle's size at this
cycle's shares; the other cycle's countries, with affinity indicators recomputed for this cycle's
nominees; and the European or non-European journalists alone (53 and
123 in 2019, 41 and 58 in 2024),
whose smaller juries add sampling variation. The bootstrap resamples the observed ballots within
region (5{,}000 replicates) without the model; it answers a related question, so agreement
checks the model's regional predictions rather than estimating the same quantity independently. A
re-vote keeps each journalist's style and precision and redraws only the model's choice noise, not
how differently the same journalist might vote.

\begin{table}[t]
\centering
\caption{Probability that the official winner wins: posterior predictive probability and 90\%
interval of $P(\mathrm{win}\mid\theta)$ across posterior draws with a confederation bonus common to
all regions and with partially pooled regional bonuses, and a bootstrap of the observed ballots
within region. 2019: Messi over van Dijk; 2024: Rodri over Vin\'icius J\'unior. Each draw's
$P(\mathrm{win}\mid\theta)$ is estimated from 100 (common) or 400 (pooled)
simulated juries, so the intervals include binomial noise with standard deviation at most
0.05 and 0.025. Under the common bonus, each probability and its interval come from separate runs. In the 2019
jury-size row the common and pooled columns agree only after rounding (0.513 and 0.506). Between-chain Monte Carlo standard errors of the probability with new journalists are
0.006 and 0.011 (common), 0.011 and
0.012 (pooled). The 2024 pooled fit is indicative. Margins under the common bonus are in Figure~\ref{fig:scenarios}.}
\label{tab:scenarios}
\small
\begin{tabular}{lrlrlr}
\toprule
Scenario & \multicolumn{2}{c}{Common bonus} & \multicolumn{2}{c}{Pooled regional bonuses} & Bootstrap \\
\cmidrule(lr){2-3}\cmidrule(lr){4-5}
 & Win & $P(\mathrm{win}\mid\theta)$ & Win & $P(\mathrm{win}\mid\theta)$ & Win \\
\midrule
\multicolumn{6}{l}{\emph{2019}} \\
Same journalists vote again & 0.57 & [0.09, 0.97] & 0.52 & [0.07, 0.93] & -- \\
New journalists, same countries & 0.55 & [0.16, 0.97] & 0.49 & [0.06, 0.93] & 0.56 \\
New journalists, affinity removed & 0.84 & [0.45, 1.00] & 0.76 & [0.28, 1.00] & -- \\
Other cycle's regional shares & 0.48 & [0.06, 0.88] & 0.40 & [0.04, 0.89] & 0.45 \\
Other cycle's jury size & 0.51 & [0.16, 0.89] & 0.51 & [0.16, 0.89] & -- \\
Other cycle's countries & 0.47 & [0.10, 0.84] & 0.43 & [0.09, 0.83] & -- \\
European journalists only & 0.06 & [0.00, 0.25] & 0.13 & [0.00, 0.45] & 0.04 \\
Non-European journalists only & 0.86 & [0.55, 1.00] & 0.77 & [0.25, 0.99] & 0.88 \\
New journalists, plurality rule & 0.55 & [0.09, 0.96] & 0.47 & [0.05, 0.90] & 0.23 \\
New journalists, Borda count & 0.55 & [0.16, 0.96] & 0.50 & [0.07, 0.92] & 0.67 \\
\midrule
\multicolumn{6}{l}{\emph{2024}} \\
Same journalists vote again & 0.77 & [0.37, 1.00] & 0.77 & [0.28, 1.00] & -- \\
New journalists, same countries & 0.70 & [0.21, 0.99] & 0.68 & [0.19, 0.99] & 0.76 \\
New journalists, affinity removed & 0.51 & [0.06, 0.97] & 0.57 & [0.06, 0.98] & -- \\
Other cycle's regional shares & 0.52 & [0.07, 0.96] & 0.55 & [0.08, 0.96] & 0.54 \\
Other cycle's jury size & 0.68 & [0.06, 1.00] & 0.72 & [0.14, 1.00] & -- \\
Other cycle's countries & 0.57 & [0.04, 1.00] & 0.58 & [0.04, 0.99] & -- \\
European journalists only & 0.96 & [0.79, 1.00] & 0.95 & [0.81, 1.00] & 1.00 \\
Non-European journalists only & 0.24 & [0.00, 0.62] & 0.24 & [0.01, 0.69] & 0.07 \\
New journalists, plurality rule & 0.81 & [0.42, 1.00] & 0.79 & [0.37, 1.00] & 0.95 \\
New journalists, Borda count & 0.62 & [0.15, 0.98] & 0.62 & [0.15, 0.96] & 0.62 \\
\bottomrule
\end{tabular}
\end{table}

\begin{figure}[!tbp]
\centering
\includegraphics[width=\textwidth]{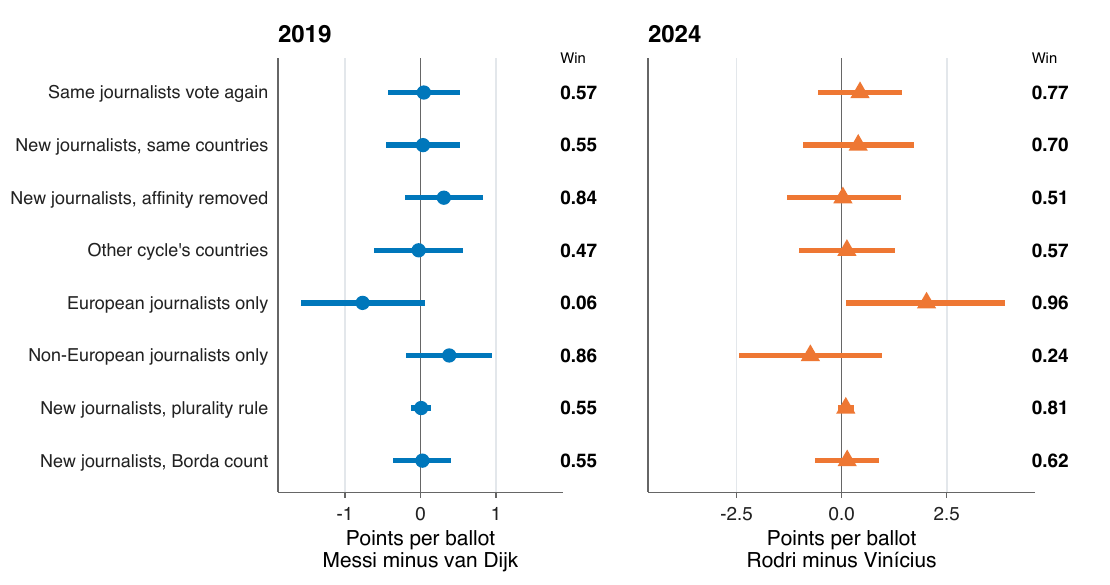}
\caption{Predicted margin of the official winner over the runner-up in official points per ballot
(the jury's margin divided by its number of journalists), with 90\% intervals, and the probability that the official winner wins (right). The official winner is
Messi in 2019 and Rodri in 2024. This figure uses the common confederation bonus to visualize
the scenario contrasts; Table~\ref{tab:scenarios} also reports partially pooled regional bonuses.
Multiplied by the jury size (176 in 2019 and 99 in 2024 for new journalists from the same
countries), the margin gives the points over the jury used in the text and Table~\ref{tab:shift}.
Removing affinity holds the other fitted parameters fixed and is not a causal intervention.}
\label{fig:scenarios}
\end{figure}

Figure~\ref{fig:scenarios} displays the corresponding margins and winning probabilities
under the common-bonus model. European and non-European juries favor different finalists in both cycles. Removing affinity helps
Messi in 2019 and Vin\'icius in 2024, and the 2024 direction is not robust to regional bonuses
(Table~\ref{tab:scenarios}).

With new journalists from the same countries, Lionel Messi wins with probability
0.55 and Rodri with 0.70
(bootstrap 0.56 and 0.76); a re-vote by the same journalists
gives 0.57 and 0.77. The model and a resampling of the ballots thus
agree that 2019 was close to a toss-up and that the 2024 result was likelier but not secure. The
intervals of $P(\mathrm{win}\mid\theta)$ are wide, 0.16 to
0.97 in 2019 and 0.21 to 0.99 in 2024 for
new journalists: the fragility reflects uncertainty about the parameters as well as the chance
composition of a jury.

\subsection{The effect of affinity}\label{sec:q1-affinity}

The coefficients $\gamma$ describe an association between
shared characteristics and rankings, given the model's account of each nominee. The effect of
removing affinity, $\Delta$, is a controlled decomposition within the fitted model: $\gamma$ is set to
zero while the tree, residuals, gate and precisions keep their posterior values. A causal effect of
journalists' preferences would require separating preference from familiarity and from regionally
valued characteristics the covariates miss. The ballots do not identify it, and we do not estimate
it. We estimate the first quantity and compute the second. $\Delta$ is not an affinity-free refit; it assumes that those parts would be unchanged if similarity played no role,
and its posterior intervals describe uncertainty within one model, not across model classes. For
each posterior draw we simulate juries of new journalists twice with common random numbers (the same
styles, precisions and Gumbel perturbations), with and without affinity (Supplement
Table~S13). In 2019 affinity lowers Lionel Messi's expected margin over
the jury from 59 points (90\% interval \ensuremath{-}3 to
123) to 11 (\ensuremath{-}42 to
68) and his probability of winning from 0.84 to
0.55; the expected favorite reverses in 0.32 of posterior
draws, and two juries that differ only in affinity elect different winners in
0.28 of pairs. In 2024 affinity raises Rodri's expected
margin from 3 to 39 points and his probability of
winning from 0.51 to 0.70; the expected favorite reverses in 0.24 of posterior draws, and
paired juries elect different winners in 0.17 of pairs. Table~\ref{tab:shift} gives the posterior of the change,
\ensuremath{-}47 points (\ensuremath{-}70 to \ensuremath{-}22) in 2019 and
37 (16 to 60) in 2024. The observed
margins of 7 and 41 points are single realizations, so the
comparison conveys the size of the effect rather than decomposing the observed result.

\begin{table}[t]
\centering
\caption{Change in the official winner's expected margin, in official points over the jury, due to
affinity, from paired simulated juries (posterior mean and 90\% interval across posterior draws).
Negative values favor the runner-up. Upper panel: the part of affinity named in each row is
removed from the primary fit; journalists from Africa, Asia-Pacific and North America share no
indicator with either finalist, and their rows (not shown) lie within five points of zero. Lower
panel: all similarity terms are removed under alternative specifications, each with its own
posterior; region-by-nominee and region-by-finalist effects are kept. Region-by-nominee effects
cannot be separated from the confederation indicator by the likelihood, so those rows depend on the prior (Supplement S6.4). The Monte Carlo standard deviation of a draw's expected change is about
1.9 points in 2019 and 2.0 in 2024.}
\label{tab:shift}
\small
\begin{tabular}{lll}
\toprule
Affinity removed & 2019 & 2024 \\
\midrule
\multicolumn{3}{l}{\emph{Part of affinity removed (common confederation bonus)}} \\
All affinity & \ensuremath{-}47 [\ensuremath{-}70, \ensuremath{-}22] & 37 [16, 60] \\
Same nation only & \ensuremath{-}1 [\ensuremath{-}3, 1] & 1 [\ensuremath{-}1, 3] \\
Same confederation only & \ensuremath{-}48 [\ensuremath{-}70, \ensuremath{-}25] & 35 [12, 60] \\
Same league country only & 0 [\ensuremath{-}1, 1] & 0 [\ensuremath{-}2, 3] \\
Journalists from Europe & \ensuremath{-}58 [\ensuremath{-}83, \ensuremath{-}33] & 46 [19, 75] \\
Journalists from South America & 10 [6, 15] & \ensuremath{-}9 [\ensuremath{-}16, \ensuremath{-}4] \\
\midrule
\multicolumn{3}{l}{\emph{All affinity removed, alternative specifications}} \\
Separate regional confederation bonuses & \ensuremath{-}40 [\ensuremath{-}68, \ensuremath{-}15] & 23 [\ensuremath{-}23, 73] \\
Partially pooled regional bonuses & \ensuremath{-}41 [\ensuremath{-}67, \ensuremath{-}17] & 24 [\ensuremath{-}22, 73] \\
Region-by-finalist effects & \ensuremath{-}50 [\ensuremath{-}71, \ensuremath{-}28] & 34 [12, 56] \\
Region-by-nominee effects, prior sd 0.25 & \ensuremath{-}50 [\ensuremath{-}76, \ensuremath{-}30] & 40 [12, 62] \\
Region-by-nominee effects, prior sd 0.5 & \ensuremath{-}47 [\ensuremath{-}74, \ensuremath{-}22] & 43 [11, 75] \\
Region-by-nominee effects, prior sd 1 & \ensuremath{-}43 [\ensuremath{-}77, \ensuremath{-}11] & 45 [4, 87] \\
One compatriot coefficient per pair & \ensuremath{-}49 [\ensuremath{-}70, \ensuremath{-}26] & 36 [14, 58] \\
\bottomrule
\end{tabular}
\end{table}

Nearly all of the effect runs through the shared confederation and the European journalists
(Table~\ref{tab:shift}). The same-nation and same-league-country terms concern few journalists
and change the margin by less than four points, and 53 journalists in 2019 and
41 in 2024 shared a confederation with the European finalist, against
10 and 10 with the South American one.

The confederation bonus may also differ by region, or stand in for
region-specific appeal of particular nominees, such as European journalists' regard for van Dijk
on grounds the season record misses. The lower panel of Table~\ref{tab:shift} refits the model with
separate or partially pooled regional bonuses and with region-by-finalist effects, one coefficient
for each region and finalist. In 2019 the bonus differs by region: with separate bonuses it is
0.60 (0.21 to 0.99) for
European and 1.91 (1.38 to
2.47) for African journalists, and the between-region standard deviation of
the pooled bonuses is 1.08 (0.49 to 1.90). The pooled bonuses predict
the 2019 ballots better than the common one, by 8.2 in expected log
predictive density (standard error 3.0, Monte Carlo error
0.3), and as well as separate bonuses
(Supplement S6). The common coefficient 0.88 thus averages a large
African and a smaller European bonus and applies the average to the European journalists, whose
ballots carry the effect on the outcome. Under the pooled bonuses the shift is
\ensuremath{-}41 (\ensuremath{-}67 to \ensuremath{-}17) points,
and removing affinity raises Lionel Messi's probability of winning from
0.49 to 0.76; with a common bonus and
region-by-finalist effects the shift is \ensuremath{-}50 (\ensuremath{-}71 to
\ensuremath{-}28). In 2024 the pooled and common bonuses predict about equally well
(difference 1.7, standard error 1.6; indicative,
as the 2024 regional fits do not meet the log-likelihood criterion). With a common bonus the shift
survives region-by-finalist effects, 34
(12 to 56) points, but not regional bonuses: with pooled
bonuses it is 24 (\ensuremath{-}22 to
73) points, and the between-region standard deviation, 0.99
(0.16 to 2.00), nearly reproduces its half-normal prior (0.06 to
1.96). Reweighting the draws to prior scales of one half and two gives shifts of
26 (\ensuremath{-}11 to 66) and
23 (\ensuremath{-}28 to 73) points
(Supplement S6.7). The 2024 ballots thus do not pin down the regional structure of
the bonus; the 2024 refits mix less well than the primary fit and are indicative.

The table also reports refits with one coefficient for every region and nominee. A journalist's
region is the confederation of their country, so the likelihood cannot separate these
effects from the confederation bonus (Supplement S6.4): those rows are a
decomposition defined by the prior, whereas the region-by-finalist refit is identified by the likelihood.

European journalists alone would have elected the European finalist in both cycles, and the other
journalists the South American one (Table~\ref{tab:scenarios}). Without affinity the 2019 split
disappears, with Lionel Messi winning with probability 0.76
in a European jury and 0.80 in a non-European one, but the 2024 split does
not (0.81 and 0.26 for
Rodri); Section~\ref{sec:q-region} asks whether the data require this split to run through voting styles.

\subsection{Jury composition and scoring rule}\label{sec:q1-design}

The 2024 jury is smaller and more European than the 2019 jury
(Section~\ref{sec:data-journalists}). With a common bonus, moving the 2019 jury to the 2024
regional shares lowers Lionel Messi's probability from 0.55 to
0.48 (bootstrap 0.45), reducing it to
99 journalists gives 0.51, and the 2024 countries give
0.47. For Rodri the corresponding probabilities are
0.52 (bootstrap 0.54), 0.68
with 176 journalists, and 0.57. Over a grid of jury sizes and
European shares (Supplement Table~S16), moving the European share from one fifth to
one half changes the official winner's probability by at least 0.30 in 2019
and 0.33 in 2024 at every size, whereas changing the size at a fixed share
changes it by at most 0.14 and 0.11, mostly because a small
jury pulls the probability toward one half. With pooled regional bonuses the share effect remains
(0.49 to 0.40 in 2019), and over the grid
the share moves the probability by at least 0.24 and
0.23, the size by at most 0.07 and
0.13. This reflects the small expected margins of these two races, not ranked juries in general, and
describes the fitted model under another jury, not the effect of the 2022 reform.

Plurality gives one point to the first name on a ballot, and Borda scoring gives $k$ points to the
first name, $k-1$ to the second and so on, with unranked nominees receiving none. With new
journalists from the same countries, plurality and Borda scoring give
Lionel Messi probabilities of 0.55 and 0.55, and
Rodri 0.81 and 0.62. On the same
simulated juries, the official rule and plurality elect different winners in
15\% of juries in 2019 and 22\% in 2024,
and the official rule and Borda in 5\% and 10\%.
The observed 2019 ballots show how the rule can decide a close race: van Dijk received more first
places than Messi (69 against 61) and was
ranked above him on 88 ballots against 86, yet
Messi won on points. For plurality the model and the bootstrap disagree in 2019
(0.55 against 0.23). The disagreement is consistent with a calibrated model: van Dijk's observed lead of 8 first places has predictive tail
probability 0.23 under new juries, while the model's expected lead for van Dijk is
\ensuremath{-}3 (\ensuremath{-}18 to 13). The bootstrap
inherits the observed first places, whereas the model estimates first-choice probabilities from
the whole ballot, so for a rule that uses first places alone the bootstrap is the more direct
guide.

\emph{Answer to Q1.} Both results were fragile under the model and under a resampling of the
ballots, and much of the fragility reflects uncertainty about the parameters. In 2019 the
confederation bonus differs by region, and under partially pooled regional bonuses the fitted model
attributes to affinity, mainly the European journalists' bonus for nominees of their own
confederation, a fall in Messi's probability of winning from 0.76 to
0.49 (from 0.84 to 0.55 with a
common bonus): the fitted model gives a near toss-up with affinity and a probable Messi win without it, and a common bonus with
region-by-finalist effects gives the same conclusion. In 2024 the model with a common
bonus attributes 37 points (16 to
60) of Rodri's expected margin to affinity, but with separate or partially
pooled regional bonuses the direction of the effect is uncertain, and the regional split remains
without affinity. In both cycles and under both bonus
specifications, the regional shares of the jury change the official winner's probability more than
its size does, and plurality and the official
rule elect different winners in 15\% and 22\% of simulated
juries.

\section{Ballots, Voting Styles and Regions}\label{sec:jury}

\subsection{Q2: what does a ballot reward?}\label{sec:q-ballot}

\begin{figure}[!tbp]
\centering
\includegraphics[width=\textwidth]{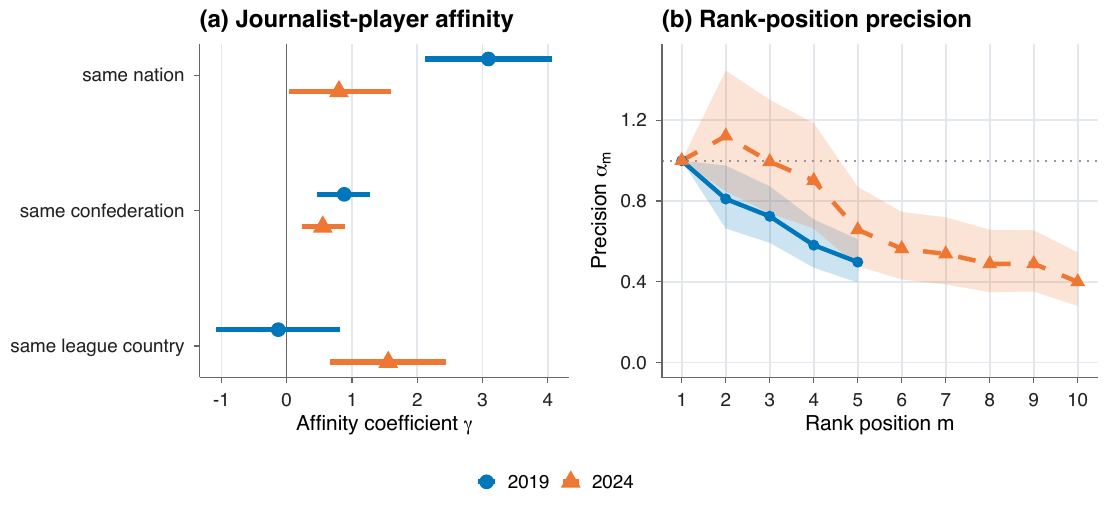}
\caption{(a) Posterior means and 90\% intervals of the affinity coefficients. (b) Posterior
means and 90\% bands of the rank-position precisions, with $\alpha_1=1$. The precisions are not
constrained to decline: in 2024 the posterior mean of $\alpha_2$ exceeds one.}
\label{fig:affinity}
\end{figure}

\emph{Affinity.} Figure~\ref{fig:affinity}a and Supplement Table~S10 report the
affinity coefficients, which we read as associations (Section~\ref{sec:q1-affinity}): how much more
readily journalists rank nominees who share their nation, confederation or league country than
nominees of equal estimated standing. In 2019 journalists favor
compatriots strongly. The same-nation coefficient is 3.09 (90\% interval
2.12 to 4.07), so for a journalist of average precision
choosing a first name, a compatriot is 21.9 times as likely to be chosen as a
nominee of equal standing from the same confederation and 53 times as
likely as one from another confederation (posterior medians). The same-confederation coefficient
is 0.88 (0.46 to 1.27), and playing in the
journalist's country adds nothing (\ensuremath{-}0.13, \ensuremath{-}1.08 to
0.82). In 2024 the same-nation coefficient falls to 0.80
(0.03 to 1.59), the same-confederation coefficient is
0.55 (0.23 to 0.90), and the
same-league-country coefficient is 1.56 (0.66 to
2.44). These estimates condition on each nominee's utility, unlike the raw
contrasts of Section~\ref{sec:data-journalists}, and agree in direction with the similarity bias
reported for the FIFA award by \citet{coupe2018} and \citet{johnsonmccarthy2022}.

The three coefficients rest on different numbers of journalists: 149
in 2019 and 73 in 2024 share a confederation with some nominee, but only
15 and 13 a nation, and
7 and 5 a league country, so the 2024
league-country coefficient describes a handful of journalists (Supplement
Table~S17 gives the counts by region). No single journalist drives the estimates:
leaving out one journalist at a time, by Pareto-smoothed importance weights, keeps the same-nation
coefficient between 2.76 and 3.35 in 2019 and the
same-league-country coefficient between 1.02 and
1.73 in 2024 (Supplement S8).

\emph{Attention down the ballot.} Choices become noisier with position
(Figure~\ref{fig:affinity}b). In 2019 the precision of the second choice is
0.81 (0.66 to 0.98) and that of the fifth
0.50 (0.40 to 0.61) relative to the first. In
2024 the second choice is as precise as the first (1.12,
0.85 to 1.45), and precision falls to
0.40 (0.28 to 0.55) at the tenth position.

\begin{figure}[!tbp]
\centering
\includegraphics[width=\textwidth]{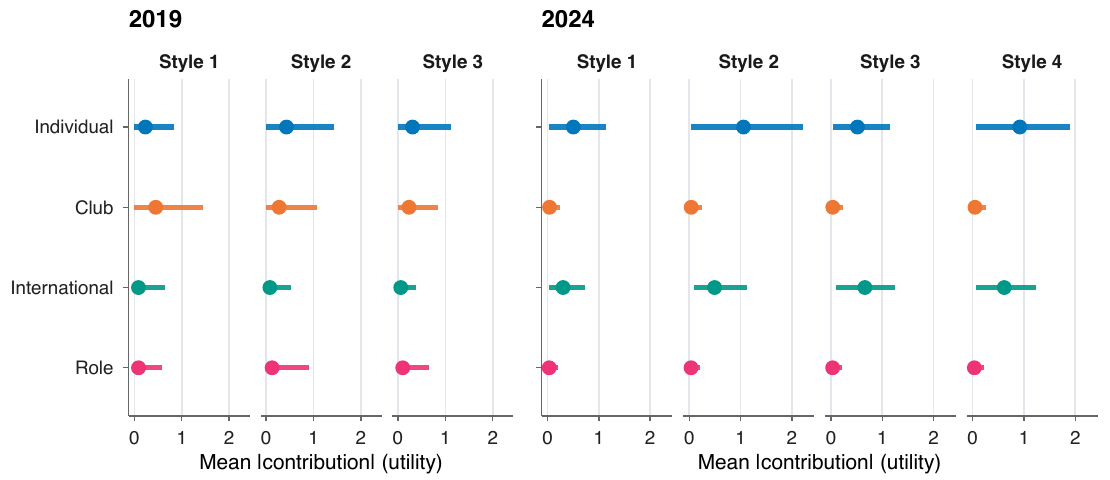}
\caption{Mean absolute baseline-Shapley contribution of each covariate group to the tree utility,
by voting style at the modal number of styles, with 90\% intervals. These are contributions
to the tree component, not shares of total nominee standing or causal effects. The 2024
style labels are not well determined and are interpreted qualitatively (Supplement S6.1).
Supplement S6.17 defines the baseline-Shapley calculation.}
\label{fig:importance}
\end{figure}

\emph{Season record and unexplained nominee effect.} The posterior favors small trees (median
3 leaves in 2019 and 5 in 2024), split most often on
the European club competition stage in 2019 and the national-team tournament result in 2024. By the
baseline-Shapley contributions of Figure~\ref{fig:importance}
\citep{sundararajan2020}, club covariates account for 30\% to
52\% of the tree component of the utility, not of the whole ballot, in 2019 and individual covariates for
27\% to 47\%, depending on the style; in
2024 club covariates contribute at most 4\%, and individual
(41\% to 65\%) and international
covariates (30\% to 53\%) share the
rest. The season record nevertheless explains a small part of the spread in standing: the ratio
of the variance of the nominee residuals to that of the style utilities is
0.92 to 0.96 across styles in 2019 and
0.76 to 0.97 in 2024. Rarely ranked nominees do not drive the
ratio: over the 10 and 17 nominees ranked on at least a
tenth of the ballots it is 0.83 to 1.05 in 2019 and
0.74 to 1.00 in 2024. The unexplained nominee effect, which contains reputation and, in 2019, the career criterion and
the autumn form missing from the covariates, therefore carries most of a nominee's standing. How
standing divides between the tree and this effect depends on the prior scale $s_b$ of the
residuals: across $s_b=0.5$, $1.0$ and $2.0$ the variance ratio ranges from 0.84 to
0.97 in 2019 and from 0.62 to 1.00 in 2024
(Supplement Table~S11). The effect carries most of the standing under each scale,
but in 2019 a smaller $s_b$ shifts the tree's share from club toward individual covariates.

\emph{Journalist precision.} Only 9 journalists in 2019 and
4 in 2024 have posterior mean precision below one half. The precision measures
agreement with the fitted styles, not the quality of a judgment.

\emph{Answer to Q2.} A nominee's standing rests mostly on what the season record does not
capture, the unexplained nominee effect, of which reputation is one component. Similarity bias is large for the journalists it concerns: the 2019
compatriot bonus of 3.09 is comparable to the gap of
3.0 between the second and the fifth nominee in the consensus
utilities, but few journalists have a compatriot among the nominees. The smaller confederation
bonus concerns most of the jury and is the part of affinity that bears on the outcome. Within the
season record, club success and individual production account for most of the tree utility in
2019, and individual and national-team performance in 2024.

\subsection{Q3: one electorate or several voting styles?}\label{sec:q-styles}

\begin{figure}[!tbp]
\centering
\includegraphics[width=\textwidth]{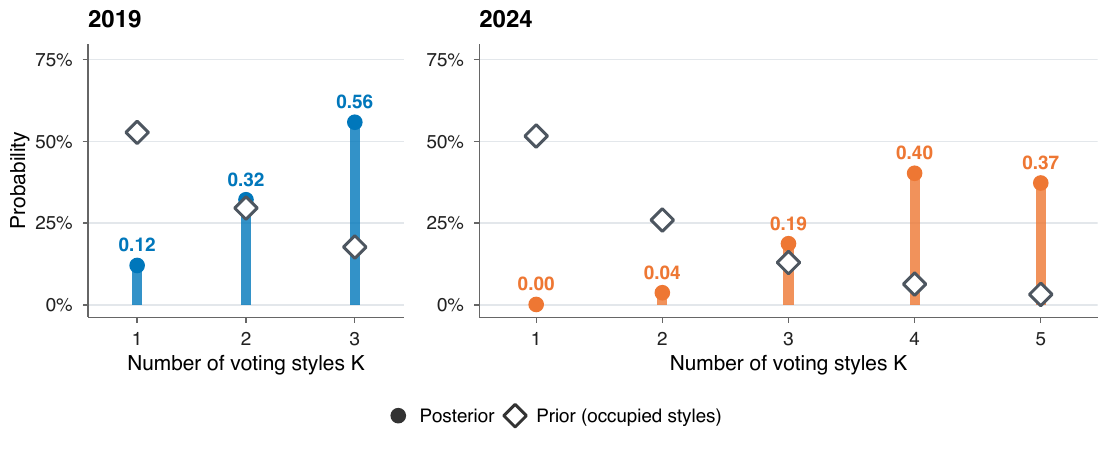}
\caption{Posterior of the number of voting styles and its prior after conditioning on every
style being occupied. The posterior puts its largest mass at or next to the bound $K_{\max}$ in
both cycles, and the allocation of most journalists to styles is uncertain
(Section~\ref{sec:q-styles}).}
\label{fig:k-posterior}
\end{figure}

Figure~\ref{fig:k-posterior} compares the posterior of the number of styles with its prior after
conditioning on occupancy. In 2019 the posterior probability of two or more styles is
0.88 against a prior probability of 0.47 (Bayes factor
8.2), and the modal number is 3, the bound. In 2024 the
posterior probability of two or more styles is 1.00, with 0.40
on four styles and 0.37 on five, the bound. The data thus indicate more than one
style but do not determine how many, and we read the styles as components that approximate
heterogeneity among journalists rather than as distinct types of voter. With $K_{\max}$ set to
2 to 5 in 2019 the posterior probability of more than one
style lies between 0.78 and 0.94, and the headline counterfactuals
barely move: the affinity shift lies between \ensuremath{-}52 and \ensuremath{-}45
points. In 2024, with $K_{\max}$ from 3 to 6, the posterior
probability of more than one style is at least 0.99 and the shift lies between
35 and 37 points (Supplement Tables~S14 and~S15).
Raising the bound moves posterior mass to more styles: the posterior mean of $K$ rises from
2.44 at $K_{\max}=3$ to 3.34 at
$K_{\max}=5$ in 2019 and from 4.11 to 4.48
at $K_{\max}=6$ in 2024, where the posterior probability of $K_{\max}$ styles is
6 and 15 times its prior probability, as
expected when styles approximate heterogeneity (Supplement S2.3).

\begin{figure}[!tbp]
\centering
\includegraphics[width=\textwidth]{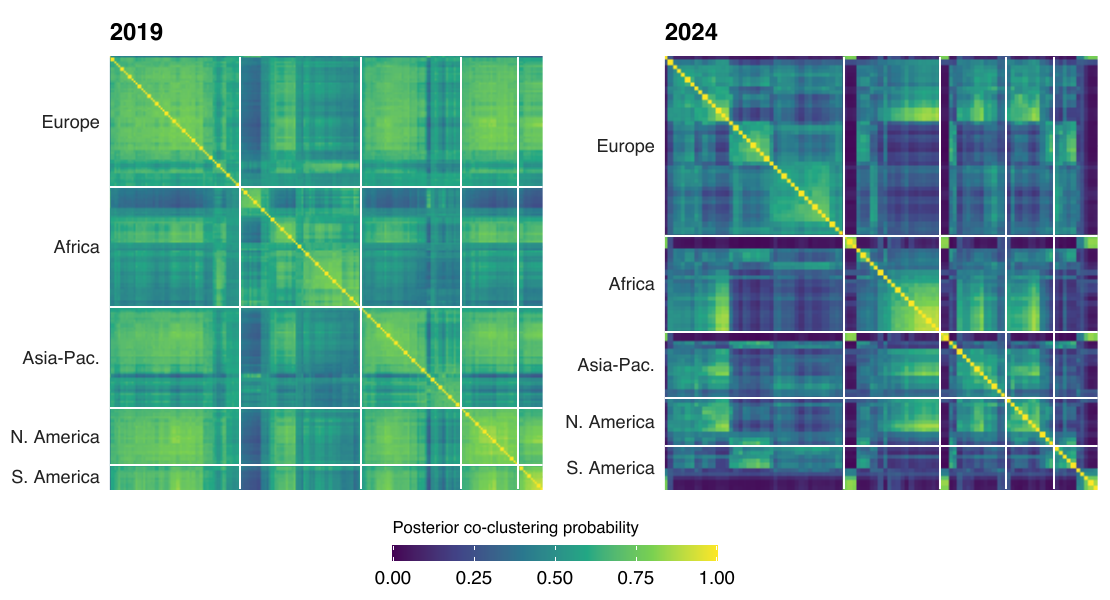}
\caption{Posterior probability that two journalists share a voting style, with journalists ordered
by region and, within region, by average-linkage clustering of these probabilities. Blocks of high
probability would indicate well-separated styles.}
\label{fig:psm}
\end{figure}

The styles are weakly separated (Figure~\ref{fig:psm}). The mean posterior entropy of a journalist's allocation is
0.98 in 2019 and 0.94 in 2024, against maxima of
1.10 and 1.39 for uniform allocations over three and four styles, so most
journalists cannot be assigned to a style with confidence. The 2019 styles differ in what they
reward: one weights club success most heavily and gives Virgil van Dijk a first-choice
probability of 0.57 against 0.20 for
Lionel Messi, and another weights individual production and reverses the order,
0.51 for Lionel Messi against
0.20 for Virgil van Dijk. In 2024 the styles divide by
finalist: in the partition that minimizes the posterior expected Binder loss, clusters totalling
52 journalists most often place Rodri first and
clusters totalling 47 place Vin\'icius J\'unior first.
In 2019 a single style with the same layers predicts nearly as well as the full model
(\ensuremath{-}1.8, standard error 2.2, in expected log predictive density),
and the posterior probability of two or more styles ranges from 0.54 to
0.99 across the priors for the journalist precisions. In 2024 a single style is worse than the best model by 11.8 (standard error 6.1) in expected
log predictive density.

\emph{Answer to Q3.} The ballots are more heterogeneous than a single ranking with affinity and
journalist precisions allows, but they do not divide the jury into well-separated styles. In 2019
the evidence for more than one style is modest and depends on the prior. In 2024 several styles
improve prediction and separate the supporters of the two finalists, but their number is not
determined. In neither cycle can most journalists be assigned to a style with confidence.

\subsection{Q4: does region predict style?}\label{sec:q-region}

Because style labels switch between draws, we measure the association between region and style by
a label-invariant contrast: the posterior probability that two journalists from the same region
share a style, minus that for two journalists from different regions. The contrast is
0.086 (90\% interval \ensuremath{-}0.009 to 0.244) in
2019 and 0.100 (0.022 to 0.229) in
2024, so journalists from the same region are somewhat more likely to share a style, clearly so
only in 2024 (Supplement Figure~S1 gives the style probabilities by region).
Region can
act through the gate or through the confederation bonus, and the posterior partly trades one for
the other: across draws, the correlation between the confederation coefficient and the mean
total-variation distance between the regions' style distributions is \ensuremath{-}0.31 in
2019 and \ensuremath{-}0.08 in 2024. Without affinity, region can affect ballots only through the gate. The European and non-European juries
of Section~\ref{sec:q-fragility} still differ without affinity in 2024 but not in 2019, so in the
fitted 2024 model the regional split runs partly through the gate. A model with a constant gate
predicts as well, however (Supplement Table~S22), and gives nearly the same outcome probabilities: under it the official
winner wins with probability 0.69 with affinity and 0.53
without in 2024 (0.56 and 0.88 in 2019; Supplement
Tables~S14 and~S15). The data therefore do not require a regional gate; it is one
decomposition the ballots allow.

\emph{Answer to Q4.} Region predicts style weakly, and the gate adds little predictive information. In the fitted 2019
model the regional difference in the outcome comes from affinity. In 2024 the fitted model routes
part of it through differences between regions in the prevalence of voting styles, but a model
without a regional gate fits the ballots as well, so this attribution is not established.

\subsection{Q5: the two cycles compared}\label{sec:q-cycles}

The nominees, ballot length, evaluation window and jury all differ between the cycles, so the
comparison is descriptive. Both races were between a European and a South American nominee, and in
both the European journalists' confederation bonus favored the European. In 2019 the model attributes to
the bonus most of the fall in Messi's probability of winning; in 2024 its effect is not established
once it may differ by region. Affinity
changed form, from a strong compatriot bonus and no league-country bonus in 2019
(3.09 and \ensuremath{-}0.13) to a weaker compatriot bonus and a league-country
bonus resting on 5 journalists in 2024 (0.80 and
1.56). The longer ballot may contribute, since ten places leave more room for
players from the home league, but two cycles cannot separate this from changes in the jury. The
criteria that separate the styles moved with the season, from club success in 2019 to individual
and national-team performance in the European Championship year, and the 2024 styles separate
the two finalists' supporters rather than criteria.

Model assessment is reported in Supplement S8. Predictive comparisons support the
performance structure, residual effects and rank-position precision, but neither predictive
checks nor style recovery establish sharply separated voting blocs. Compatriot votes remain
under-predicted, and the 2024 comparisons inherit the convergence limitations of their fits.

\section{Discussion}\label{sec:discussion}

We develop a Bayesian framework for partial rankings that combines a shared soft tree,
an unknown number of voting styles, region-dependent mixing probabilities and
journalist--nominee affinity. The common performance basis makes styles comparable,
while joint posterior inference propagates uncertainty to award-outcome comparisons.
Identification and consistency hold under stated conditions. In the Ballon d'Or data, the fitted
model attributes much of the closeness of the 2019 race to the European journalists'
confederation bonus, leaves the direction of the 2024 affinity effect unsettled, and finds that
regional representation changes winning probabilities more than jury size, although the ballots
do not split into sharply separated voting blocs.

Several limitations remain. Affinity measures association, not a causal preference effect,
and can reflect familiarity or omitted regional criteria. The two cycles differ in
nominees, electorates and evaluation windows, so they do not identify the effect of the
2022 reform. Each has only 30 nominee records, with imperfect performance measurements
and weakly separated styles. Identification relies on an unverified expert-mixture
condition, and style-count consistency requires a distinct-style prior restriction not
imposed in the fitted model. Computational diagnostics also limit the 2024 style summaries
and auxiliary comparisons (Supplement S6.1). The outcome comparisons are model-based
predictions under specified changes, not causal claims about alternative juries.

Future work could extend the framework longitudinally, allowing performance relationships,
voting styles and regional representation to evolve across awards while accounting for
changing nominees and rules. A second direction is conformal inference for prediction,
using ranking-based conformity scores to construct prediction sets for future ballots,
nominee ranks or award winners. Establishing coverage would require calibration schemes
suited to dependent voters, temporal change and varying candidate sets; the present
posterior intervals do not provide such guarantees. Richer affinity measures and
hierarchical country effects could further distinguish national from regional variation.

\section*{Supplementary material}
Data construction and corrections (S1); the theory summarized in
Section~\ref{sec:theory} and its proofs (S2); the priors and the sampler's moves
and acceptance ratios (S3); validation of the sampler (S4); the simulation study
(S5); additional results (S6); sensitivity analyses
(S7); model assessment (S8).
\subsection*{Code and data}
A zip archive with the ballots, the nominee and journalist data, and the code that builds the data,
fits the models and writes every number, table and figure of the article into its LaTeX sources,
with the stored result summaries but not the posterior draws, which refitting rebuilds (run times
in Supplement S3.2).
The licenses of the Opta-derived league measures may restrict redistribution, so for these columns
the archive names the source tables (with URLs) and gives the player and season identifiers and
the merging code in place of the values.

\bibliographystyle{plainnat}
\bibliography{refs}

@misc{pearce2026benter,
  author = {Pearce, Michael},
  title = {Efficient {Bayesian} Inference for {Benter} Models on Ranked Data},
  year = {2026},
  eprint = {2608.23825},
  archivePrefix = {arXiv},
  primaryClass = {stat.ME},
  doi = {10.48550/arXiv.2608.23825}
}

@article{iacopini2026bart,
  author = {Iacopini, Matteo and O'Neill, Eoghan and Rossini, Luca},
  title = {Static and Dynamic {BART} for Rank-Order Data},
  journal = {Journal of Business \& Economic Statistics},
  year = {2026},
  doi = {10.1080/07350015.2025.2604128},
  note = {Published online 13 April 2026}
}

@article{deng2014bard,
  author = {Deng, Ke and Han, Simeng and Li, Kate J. and Liu, Jun S.},
  title = {Bayesian Aggregation of Order-Based Rank Data},
  journal = {Journal of the American Statistical Association},
  year = {2014},
  volume = {109},
  number = {507},
  pages = {1023--1039},
  doi = {10.1080/01621459.2013.878660}
}

@article{li2022barc,
  author = {Li, Xinran and Yi, Dingdong and Liu, Jun S.},
  title = {Bayesian Analysis of Rank Data with Covariates and Heterogeneous Rankers},
  journal = {Statistical Science},
  year = {2022},
  volume = {37},
  number = {1},
  pages = {1--23},
  doi = {10.1214/20-STS818}
}

@article{plackett1975,
  author   = {Plackett, R. L.},
  title    = {The Analysis of Permutations},
  journal  = {Applied Statistics},
  volume   = {24},
  number   = {2},
  pages    = {193--202},
  year     = {1975},
  doi      = {10.2307/2346567}
}

@article{chipman1998,
  author   = {Chipman, Hugh A. and George, Edward I. and McCulloch, Robert E.},
  title    = {{Bayesian} {CART} Model Search},
  journal  = {Journal of the American Statistical Association},
  volume   = {93},
  number   = {443},
  pages    = {935--948},
  year     = {1998},
  doi      = {10.1080/01621459.1998.10473750}
}

@article{miller2018,
  author   = {Miller, Jeffrey W. and Harrison, Matthew T.},
  title    = {Mixture Models With a Prior on the Number of Components},
  journal  = {Journal of the American Statistical Association},
  volume   = {113},
  number   = {521},
  pages    = {340--356},
  year     = {2018},
  doi      = {10.1080/01621459.2016.1255636}
}

@article{fruhwirth2021,
  author   = {Fr{\"u}hwirth-Schnatter, Sylvia and Malsiner-Walli, Gertraud and Gr{\"u}n, Bettina},
  title    = {Generalized Mixtures of Finite Mixtures and Telescoping Sampling},
  journal  = {Bayesian Analysis},
  volume   = {16},
  number   = {4},
  pages    = {1279--1307},
  year     = {2021},
  doi      = {10.1214/21-ba1294}
}

@article{linero2018,
  author   = {Linero, Antonio R. and Yang, Yun},
  title    = {{Bayesian} Regression Tree Ensembles that Adapt to Smoothness and Sparsity},
  journal  = {Journal of the Royal Statistical Society Series B: Statistical Methodology},
  volume   = {80},
  number   = {5},
  pages    = {1087--1110},
  year     = {2018},
  doi      = {10.1111/rssb.12293}
}

@article{gormley2008aoas,
  author   = {Gormley, Isobel Claire and Murphy, Thomas Brendan},
  title    = {A mixture of experts model for rank data with applications in election studies},
  journal  = {The Annals of Applied Statistics},
  volume   = {2},
  number   = {4},
  pages    = {1452--1477},
  year     = {2008},
  doi      = {10.1214/08-aoas178}
}

@incollection{benter2008,
  author   = {Benter, William},
  title    = {Computer Based Horse Race Handicapping and Wagering Systems: A Report},
  booktitle= {Efficiency of Racetrack Betting Markets},
  editor   = {Hausch, Donald B. and Lo, Victor S. Y. and Ziemba, William T.},
  pages    = {183--198},
  publisher= {World Scientific},
  address  = {Singapore},
  note     = {Reprint of the 1994 Academic Press edition},
  year     = {2008},
  doi      = {10.1142/9789812819192_0019}
}

@article{johnson2020,
  author   = {Johnson, S. R. and Henderson, D. A. and Boys, R. J.},
  title    = {Revealing Subgroup Structure in Ranked Data Using a {Bayesian} {WAND}},
  journal  = {Journal of the American Statistical Association},
  volume   = {115},
  number   = {532},
  pages    = {1888--1901},
  year     = {2020},
  doi      = {10.1080/01621459.2019.1665528}
}

@article{mollica2017,
  author   = {Mollica, Cristina and Tardella, Luca},
  title    = {{Bayesian} {Plackett--Luce} Mixture Models for Partially Ranked Data},
  journal  = {Psychometrika},
  volume   = {82},
  number   = {2},
  pages    = {442--458},
  year     = {2017},
  doi      = {10.1007/s11336-016-9530-0}
}

@article{jacobs1991,
  author   = {Jacobs, Robert A. and Jordan, Michael I. and Nowlan, Steven J. and Hinton, Geoffrey E.},
  title    = {Adaptive Mixtures of Local Experts},
  journal  = {Neural Computation},
  volume   = {3},
  number   = {1},
  pages    = {79--87},
  year     = {1991},
  doi      = {10.1162/neco.1991.3.1.79}
}

@article{kamakura1989,
  author   = {Kamakura, Wagner A. and Russell, Gary J.},
  title    = {A Probabilistic Choice Model for Market Segmentation and Elasticity Structure},
  journal  = {Journal of Marketing Research},
  volume   = {26},
  number   = {4},
  pages    = {379--390},
  year     = {1989},
  doi      = {10.1177/002224378902600401}
}

@article{gupta1994,
  author   = {Gupta, Sachin and Chintagunta, Pradeep K.},
  title    = {On Using Demographic Variables to Determine Segment Membership in Logit Mixture Models},
  journal  = {Journal of Marketing Research},
  volume   = {31},
  number   = {1},
  pages    = {128--136},
  year     = {1994},
  doi      = {10.1177/002224379403100111}
}

@article{coupe2018,
  author   = {Coup{\'e}, Tom and Gergaud, Olivier and Noury, Abdul},
  title    = {Biases and Strategic Behaviour in Performance Evaluation: The Case of the {FIFA's} best soccer player award},
  journal  = {Oxford Bulletin of Economics and Statistics},
  volume   = {80},
  number   = {2},
  pages    = {358--379},
  year     = {2018},
  doi      = {10.1111/obes.12201}
}

@article{johnsonmccarthy2022,
  author   = {Johnson, Michael R. and McCarthy, Ian P.},
  title    = {Cultural similarity and impartiality on voting bias: The case of {FIFA's} {World's} {Best} {Male} {Football} {Player} {Award}},
  journal  = {PLOS ONE},
  volume   = {17},
  number   = {7},
  pages    = {e0270546},
  year     = {2022},
  doi      = {10.1371/journal.pone.0270546}
}

@article{kopkin2022,
  author   = {Kopkin, Nolan and Roberts, Andrew},
  title    = {Biases in elections with well-informed voters: Evidence from public voting for football awards},
  journal  = {Social Science Quarterly},
  volume   = {103},
  number   = {7},
  pages    = {1551--1571},
  year     = {2022},
  doi      = {10.1111/ssqu.13216}
}

@article{zitzewitz2006,
  author   = {Zitzewitz, Eric},
  title    = {Nationalism in Winter Sports Judging and Its Lessons for Organizational Decision Making},
  journal  = {Journal of Economics \& Management Strategy},
  volume   = {15},
  number   = {1},
  pages    = {67--99},
  year     = {2006},
  doi      = {10.1111/j.1530-9134.2006.00092.x}
}

@article{emerson2009,
  author   = {Emerson, John W. and Seltzer, Miki and Lin, David},
  title    = {Assessing Judging Bias: An Example From the 2000 {Olympic} Games},
  journal  = {The American Statistician},
  volume   = {63},
  number   = {2},
  pages    = {124--131},
  year     = {2009},
  doi      = {10.1198/tast.2009.0026}
}

@article{ginsburgh2008,
  author   = {Ginsburgh, Victor and Noury, Abdul G.},
  title    = {The {Eurovision} {Song} {Contest}. Is voting political or cultural?},
  journal  = {European Journal of Political Economy},
  volume   = {24},
  number   = {1},
  pages    = {41--52},
  year     = {2008},
  doi      = {10.1016/j.ejpoleco.2007.05.004}
}

@article{spierdijk2009,
  author   = {Spierdijk, Laura and Vellekoop, Michel},
  title    = {The structure of bias in peer voting systems: lessons from the {Eurovision} {Song} {Contest}},
  journal  = {Empirical Economics},
  volume   = {36},
  number   = {2},
  pages    = {403--425},
  year     = {2009},
  doi      = {10.1007/s00181-008-0202-5}
}

@article{hausman1987,
  author   = {Hausman, Jerry A. and Ruud, Paul A.},
  title    = {Specifying and Testing Econometric Models for Rank-Ordered Data},
  journal  = {Journal of Econometrics},
  volume   = {34},
  number   = {1--2},
  pages    = {83--104},
  year     = {1987},
  doi      = {10.1016/0304-4076(87)90068-6}
}

@article{fok2012,
  author   = {Fok, Dennis and Paap, Richard and van Dijk, Bram},
  title    = {A Rank-Ordered Logit Model with Unobserved Heterogeneity in Ranking Capabilities},
  journal  = {Journal of Applied Econometrics},
  volume   = {27},
  number   = {5},
  pages    = {831--846},
  year     = {2012},
  doi      = {10.1002/jae.1223}
}

@article{caron2014,
  author   = {Caron, Fran{\c{c}}ois and Teh, Yee Whye and Murphy, Thomas Brendan},
  title    = {{Bayesian} Nonparametric {Plackett--Luce} Models for the Analysis of Preferences for College Degree Programmes},
  journal  = {The Annals of Applied Statistics},
  volume   = {8},
  number   = {2},
  pages    = {1145--1181},
  year     = {2014},
  doi      = {10.1214/14-aoas717}
}

@article{li2017,
  author   = {Li, Danielle},
  title    = {Expertise versus Bias in Evaluation: Evidence from the {NIH}},
  journal  = {American Economic Journal: Applied Economics},
  volume   = {9},
  number   = {2},
  pages    = {60--92},
  year     = {2017},
  doi      = {10.1257/app.20150421}
}

@misc{tnt2022,
  author       = {{TNT Sports}},
  title        = {Ballon d'{Or} 2022: How Does the Format for the {Ballon d'Or} Work? Changes Made to Football's Prestigious Award},
  howpublished = {\url{https://www.tntsports.co.uk/football/ballon-d-or/2021/ballon-dor-2022-how-does-the-format-for-the-ballon-dor-work-changes-made-to-footballs-prestigious-aw_sto9181603/story.shtml}},
  note         = {Updated 12 October 2022; accessed 11 October 2026},
  year         = {2022}
}

@book{luce1959,
  author    = {Luce, R. Duncan},
  title     = {Individual Choice Behavior: A Theoretical Analysis},
  publisher = {Wiley},
  address   = {New York},
  year      = {1959}
}

@inproceedings{zhao2019,
  author    = {Zhao, Zhibing and Xia, Lirong},
  title     = {Learning Mixtures of {Plackett--Luce} Models from Structured Partial Orders},
  booktitle = {Advances in Neural Information Processing Systems 32},
  pages     = {10143--10153},
  year      = {2019}
}

@inproceedings{cai2021,
  author    = {Cai, Diana and Campbell, Trevor and Broderick, Tamara},
  title     = {Finite Mixture Models Do Not Reliably Learn the Number of Components},
  booktitle = {Proceedings of the 38th International Conference on Machine Learning},
  series    = {Proceedings of Machine Learning Research},
  volume    = {139},
  pages     = {1158--1169},
  publisher = {PMLR},
  year      = {2021}
}

@inproceedings{irsoy2012,
  author    = {Irsoy, Ozan and Yildiz, Olcay Taner and Alpaydin, Ethem},
  title     = {Soft Decision Trees},
  booktitle = {Proceedings of the 21st International Conference on Pattern Recognition},
  pages     = {1819--1822},
  year      = {2012}
}

@inproceedings{sundararajan2020,
  author    = {Sundararajan, Mukund and Najmi, Amir},
  title     = {The Many {Shapley} Values for Model Explanation},
  booktitle = {Proceedings of the 37th International Conference on Machine Learning},
  series    = {Proceedings of Machine Learning Research},
  volume    = {119},
  pages     = {9269--9278},
  publisher = {PMLR},
  year      = {2020}
}

@misc{afp2022,
  author       = {{Agence France-Presse}},
  title        = {Ballon d'{Or} to Revamp Format from Next Season},
  howpublished = {The 42, \url{https://www.the42.ie/ballon-dor-format-5708303-Mar2022/}},
  note         = {Published 11 March 2022; accessed 10 October 2026},
  year         = {2022}
}

@misc{baldwin2019,
  author = {Baldwin, Alan},
  title = {Van {Dijk} Deserved to Win {Ballon d'Or}, Says {Klopp}},
  year = {2019},
  howpublished = {Reuters, republished by Euronews},
  note = {3 December 2019},
  url = {https://www.euronews.com/2019/12/03/van-dijk-deserved-to-win-ballon-dor-says-klopp}
}

@misc{hofmanfaez2024,
  author = {Hofman, Gustavo and F{\'a}ez, Rodrigo},
  title = {Vin{\'i}cius {J{\'u}nior}, {Real Madrid} to Skip {Ballon d'Or} Awards},
  year = {2024},
  howpublished = {ESPN},
  note = {28 October 2024},
  url = {https://global.espn.com/football/story/_/id/42052014/child/child/Message}
}

@article{santi2026btsbm,
  author = {Santi, Lapo and Friel, Nial},
  title = {The {Bradley--Terry} stochastic block model},
  journal = {The Annals of Applied Statistics},
  year = {2026},
  volume = {20},
  number = {2},
  pages = {1766--1787},
  doi = {10.1214/26-AOAS2193}
}

@misc{santi2026plblocks,
  author = {Santi, Lapo and Friel, Nial and Vitelli, Valeria},
  title = {Bayesian {Plackett--Luce} Latent Block Models for Ranked Data},
  year = {2026},
  eprint = {2607.26949},
  archivePrefix = {arXiv},
  primaryClass = {stat.ME},
  doi = {10.48550/arXiv.2607.26949},
  url = {https://arxiv.org/abs/2607.26949}
}

@article{pearce2025preferences,
  author = {Pearce, Michael and Erosheva, Elena A.},
  title = {Modeling Preferences: A {Bayesian} Mixture of Finite Mixtures for Rankings and Ratings},
  journal = {Journal of the American Statistical Association},
  year = {2025},
  volume = {120},
  number = {551},
  pages = {1621--1632},
  doi = {10.1080/01621459.2024.2444700}
}

@misc{troude2026criteria,
  author = {Troude, Th{\'e}o},
  title = {What Do You Really Need to Do to Win the {Ballon d'Or}?},
  year = {2026},
  howpublished = {Ballon d'Or, France Football},
  url = {https://ballondor.com/news/posts/what-do-you-really-need-to-do-to-win-the-ballon-dor}
}

@misc{uefa2024rules,
  author = {{UEFA}},
  title = {{Ballon d'Or}: How Are the Winners Decided?},
  year = {2024},
  note = {28 October 2024},
  url = {https://www.uefa.com/ballondor/news/0292-1c2705cb8fd5-8f2dcb111a8a-1000--ballon-d-or-how-are-the-winners-decided/}
}

@misc{uefa2024,
  author = {{UEFA}},
  title = {Rodri Wins 2024 Men's {Ballon d'Or}},
  year = {2024},
  note = {28 October 2024},
  url = {https://www.uefa.com/uefachampionsleague/news/0292-1c33aa864c69-d47a84004f4d-1000--rodri-wins-2024-men-s-ballon-d-or/}
}

@misc{hu2026supp,
  author = {Hu, Guanyu and Wang, Xiao},
  title = {Supplement to ``{Close Races and Similarity Bias: A Bayesian Analysis of the Ranked Ballots for the Ballon d'Or}''},
  year = {2026}
}

@inproceedings{schafer2015,
  author    = {Sch{\"a}fer, Dirk and H{\"u}llermeier, Eyke},
  title     = {Dyad Ranking Using a Bilinear {Plackett--Luce} Model},
  booktitle = {Machine Learning and Knowledge Discovery in Databases (ECML PKDD 2015)},
  series    = {Lecture Notes in Computer Science},
  pages     = {227--242},
  publisher = {Springer},
  address   = {Cham},
  year      = {2015},
  doi       = {10.1007/978-3-319-23525-7_14}
}

@article{schafer2018,
  author  = {Sch{\"a}fer, Dirk and H{\"u}llermeier, Eyke},
  title   = {Dyad Ranking Using {Plackett--Luce} Models Based on Joint Feature Representations},
  journal = {Machine Learning},
  volume  = {107},
  number  = {5},
  pages   = {903--941},
  year    = {2018},
  doi     = {10.1007/s10994-017-5694-9}
}

\end{document}